\documentclass[a4paper,11pt]{article}
\usepackage[latin1]{inputenc}
\usepackage[T1]{fontenc}
\usepackage{graphicx}
\usepackage{amsfonts}
\usepackage{array}
\usepackage{textcomp}
\usepackage{amssymb}
\usepackage{xcolor}
\usepackage{latexsym}
\usepackage{amsmath}

\newcommand{\ittem}{\item[$\bullet$]}
\newcommand{\Sp}{\textrm{Spin}}
\newcommand{\vt}{\vartheta}
\newcommand{\Z}{\mathbb Z}

\newcommand{\ket}{\rangle}
\newcommand{\Th}[2]{\vt\left[^{#1}_{#2}\right]}
\newcommand{\Thbar}[2]{\bar\vt\left[^{#1}_{#2}\right]}
\newcommand{\ab}{\bar a}
\newcommand{\bb}{\bar b}
\newcommand{\gb}{\bar \gamma}
\newcommand{\db}{\bar \delta}

\newcommand{\cN}{{\cal N}}
\newcommand{\cB}{{\cal B}}
\newcommand{\be}[0]{\begin{equation}}
\newcommand{\ee}[0]{\end{equation}}
\newcommand{\bi}[0]{\begin{itemize}}
\newcommand{\ei}[0]{\end{itemize}}
\newcommand{\ba}[0]{\begin{eqnarray}}
\newcommand{\ea}[0]{\end{eqnarray}}
\newcommand{\bal}{\begin{align}}
\newcommand{\eal}{\end{align}}

\newcommand{\non}{\nonumber}

\numberwithin{equation}{section}

\begin{document}

\begin{titlepage}
\begin{flushright}
LPTENS--08/33\\
LTH 797\\
July 2008
\end{flushright}

\vspace{1cm}

\begin{center}

{\LARGE\bf Spinor-Vector Duality in Heterotic SUSY Vacua}


\vspace{0.7cm}

{\large Tristan Catelin-Jullien$^*$, Alon E. Faraggi$^\dag$, Costas Kounnas$^*$\\and John
Rizos$^\ddag$}

\vspace{0.7cm}

 $^*$~Laboratoire de Physique Th\'eorique\footnote{Unit\'e mixte
du CNRS et de l'Ecole Normale Sup\'erieure associ\'ee \`a l'universit\'e Pierre et Marie Curie
(Paris 6), UMR 8549.\\}, Ecole
Normale Sup\'erieure,\\
24 rue Lhomond, F--75231 Paris cedex 05, France.\\
{\em catelin@lpt.ens.fr, costas.kounnas@lpt.ens.fr}

\vspace{0.3cm}

$^\dag$~Department of
Mathematical Sciences, University of Liverpool, Liverpool L69 7ZL, UK.\\
{\em  Alon.Faraggi@liv.ac.uk}

 \vspace{0.3cm}

 $^\ddag$~Department of Physics,
University of Ioannina, GR45110 Ioannina, Greece.\\
{\em  irizos@uoi.gr}

\bigskip

\end{center}

\vspace{0.7cm}

\begin{abstract}

We elaborate on the recently discovered spinor-vector duality in realistic free fermionic heterotic
vacua. We emphasize the interpretation of the freely-acting orbifolds carried out on the six
internal dimensions as coordinate-dependent compactifications; they play a central role in the
duality, especially because of their ability to break the right-moving superconformal algebra of
the space-time supersymmetric heterotic vacua. These considerations lead to a simple and intuitive
proof of the spinor-vector duality, and to the formulation of explicit rules to find the dual of a
given model. We discuss the interest of such a duality, notably
concerning the structure of the space of vacua of superstring theory.\\

\end{abstract}


\end{titlepage}

\section{Introduction}

Heterotic string theory \cite{Gross} is a preferred candidate to build realistic string theories.
Indeed, its structure allows a large variety of gauge groups, derived from the breaking of the
original $SO(32)$ or $E_8\times E_8$ 10-dimensional gauge group upon compactification
\cite{CalabiYau}. These groups include usual grand unification groups such as $SO(10)$ or $SU(5)$,
usually arising from the breaking of the $E_6$ gauge group present in a $N=(2,2)$ Calabi-Yau
compactification of heterotic string theories.

\bigskip

One expects a realistic theory to have $\cN=1$ (which is further spontaneously broken)
four-dimensional supersymmetry. In our framework, this is achieved by compactifying the six
internal dimensions on a $T^6/\Z_2\times \Z_2$ orbifold. This procedure initially breaks
supersymmetry from $\cN=4$ to $\cN=1$. The last breaking $\cN=1 \to \cN=0$ is assumed to be
realized either by non-perturbative phenomena or by (geometric or non-geometric) fluxes
\cite{StringySS}. The $T^6/\Z_2\times \Z_2$ orbifold framework  also has the advantage to have
three $\cN=2$ twisted sectors, which can lead naturally to a realization of models with three
generations \cite{FKR} -- \cite{z2z2models2}.

\bigskip

The models we are going to be interested in are built using the so-called fermionic construction
\cite{FF}, where the Weyl anomaly is cancelled by inclusion of free fermionic degrees of freedom on
the world-sheet. Over the years, several string-derived realistic models have been constructed
using this formalism \cite{Realistic}. It is known \cite{z2z2models1,z2z2models2} that such models
reproduce a wide variety of compactifications, toroidal or more generally Calabi-Yau, at special
points of their moduli space. A particular model is specified by a basis of sets of fermions, or
more precisely by summation over a set of spin structures authorized for the fermions. In this
procedure, standard $\Z_2$ freely-acting and non-freely acting orbifolds are encoded in a very
natural way, which arises from the properties of fermionization when the internal manifold is at
the extended symmetry point, referred to as the \emph{fermionic point}. Placing ourselves at this
specific point of the moduli space of the theory is not very restrictive : indeed, if one chooses
to deform these models in order to move away from this point, the form of the twisted sectors, and
therefore the chiral matter content of the model, is unchanged as these sectors are insensitive to
the geometry of the compactification manifold \cite{FKR,z2z2models2,FKNR}. The $T^6/\Z_2\times
\Z_2$ orbifold breaking the supersymmetry to $\cN=1$ is realized by means of the introduction of
two sets of fermions, that we will call $b_1$ and $b_2$. We finally have to specify the value of
various discrete torsion coefficients, defining the action of the generalized GSO projections
present in the construction; this specification, among other things, encodes the precise effect of
all the orbifoldings that have been introduced.

\bigskip

In this paper, we will focus on a duality that has been pointed out in a recent work \cite{FKR},
where several properties of all possible heterotic $\Z_2 \times \Z_2$ models have been detailed, by
means of a computerized statistical study of their massless spectra. This study has been restricted
to a subclass of models closely resembling the usual three generation realistic string models,
where the gauge group yielded by the free fermions include a factor $SO(10)$. This duality
exchanges, within the three twisted sectors of the orbifold, the number of vectorial
representations of $SO(10)$ with the number of spinorial plus anti-spinorial representations of
$SO(10)$. Starting from obviously self-dual cases, namely the cases where the $SO(10)$ gauge is
extended to $E_6$, which can be linked to the usual $N=(2,2)$ compactifications on Calabi-Yau
surfaces, we will be able to project out some of the representations of $SO(10)$ by suitable
freely-acting orbifolds, therefore explicitly creating dual pairs of models in a straightforward
way. We will be able to construct the dual model of some generic model, which will prove the
duality. As noted in previous work, this duality is realized internally in each twisted sector.
Consequentially, the duality has been shown to hold in $\cN=2$ theories as well (as $\cN=2$
supersymmetry is conserved in each of the twisted sectors). The mechanism of the proof can be
adapted in a straightforward way to this case.

\bigskip

The main ingredient of the construction will be to  consider the effect of freely-acting orbifolds.
These orbifolds, when carried out in the simplest way, correspond to the modding out of a
half-shift symmetry $X \to X+\pi R$ on an internal boson $X$. In this case, the generated twisted
sectors are massive; without further hypotheses, the mass shift does not depend on the various
representations to which the states belong. However, in a particular framework, the freely-acting
orbifold can break a symmetry by lifting the mass degeneracy between the symmetry partners. This
happens if, in addition to the translation, we consider modding out a parity operator,
discriminating states having different charges under a symmetry group. As a result, states with
different charges will undergo different mass shifts, leading to a \emph{spontaneous} breaking of
symmetry. This mechanism is the stringy generalization \cite{StringySS} of the field-theoretic
Scherk-Schwarz compactification \cite{SS} ; it can be used to spontaneously break supersymmetry,
when the parity operator is chosen to be the space-time helicity of the string state
\cite{StringySS}. More generally, various patterns of spontaneous SUSY breaking are obtained by
choosing an arbitrary $R$-symmetry charge (see for example \cite{CJKPT} for a recent cosmological
application of these constructions).

This enables us also to break an internal superconformal algebra, relating vectorial and spinorial
representations of some gauge group of the theory. The current transforming the spinorial
representation into the vectorial one and vice-versa is part of the right hand side of the
$N=(2,2)$ superconformal algebra present in the model in the case of an unbroken $E_6$. By doing a
Scherk-Schwarz compactification of an internal direction coupled to the helicity associated to the
different representations of the gauge group, one is then able to break this superconformal
symmetry, discriminating vectorial and spinorial representations by creating a mass gap.

\bigskip

In the first part of this paper, we will review the free fermionic setup used to construct the
class of models we will be interested in. Then we will detail how one can implement freely-acting
orbifolds with the sets we introduced, how these freely-acting orbifolds can be used for the
spontaneously breaking of some symmetry, and how it can, in our case, lift the mass degeneracy
between the spinorial/anti-spinorial representations of $SO(10)$ and the vectorial representations
of $SO(10)$. In a third part, we will focus on one twisted plane (that is, one family of twisted
sectors) of the theory. We will start by considering one specific model in the first twisted plane,
and detail its massless spectrum. Then, we will enunciate the rules to construct the $(S_t
\leftrightarrow V)$-dual of a model, and apply them on the model we just constructed. We will also
give some tools to perform this duality directly on the partition function of the theory. Finally,
we will conclude by some remarks on the significance of this duality, especially regarding the
structure of the vacua of $\cN=1$ heterotic string theories.

\section{Free fermionic construction}

\subsection{$\cN=1$ and $\cN=2$ parity set basis and partition function}

Starting for a four-dimensional superstring theory made out of free fermions \cite{FF}, the 20
left-moving fermions are noted, following references \cite{FKNR,Nooij}

\be \{\psi^\mu,\chi^{1\dots 6},y^{1\dots 6},\omega^{1\dots 6}\} \ee and the 44 right-moving ones
\be \{\bar y^{1\dots 6},\bar\omega^{1\dots 6},\bar \psi^{1\dots 5},\bar \eta^{1\dots
3},\bar\phi^{1\dots 8}\} \ee where the $\bar\psi$'s, $\bar\eta$'s and $\bar\phi$'s are
\emph{complex} fermions. These notations fixed, we are considering the sets \be \non F=
\{\psi^\mu,\chi^{1\dots 6},y^{1\dots 6},\omega^{1\dots 6}~|~\bar y^{1\dots 6},\bar\omega^{1\dots
6},\bar \psi^{1\dots 5},\bar \eta^{1\dots 3},\bar\phi^{1\dots 8}\};\ee

\be \non S= \{\psi^\mu,\chi^{1\dots 6}\};~~~~~~e_i= \{y^i,\omega^i~|~\bar
y^i,\bar\omega^i\},~~\left[\,i=1\dots 6\,\right];\ee

\be \label{full_basis} b_1= \{\chi^{3\dots 6},y^{3\dots 6}~|~\bar y^{3\dots 6},\bar \psi^{1\dots
5},\bar \eta^1\};\ee

\be b_2= \non \{\chi^{1,2,5,6},y^{1,2,5,6}~|~\bar y^{1,2,5,6},\bar \psi^{1\dots 5},\bar
\eta^2\};\ee

\be \non z_1= \{\bar\phi^{1\dots 4}\};~~~~~~ z_2= \{\bar\phi^{5\dots 8}\}.\ee Noting additively the
usual composition law of the free fermionic formalism, we will use that

\be x = \{\bar\psi^{1\dots 5},\bar\eta^{1,2,3}\} = F+S+\sum_i e_i+z_1+z_2 \ee and \be b_3 =
b_1+b_2+x = \{\chi^{1\dots 4},y^{1\dots 4}~|~\bar y^{1\dots 4},\bar \psi^{1\dots 5},\bar \eta^3\}
\ee are part of the vacua of the theory. Note that the case of a $\cN=2$ theory is treated by
considering the previous set, amputated of $b_2$. This has the effect of considering a
$T^4/\Z_2\times T^2$ orbifold instead of a $T^6/\Z_2 \times \Z_2$. The duality also holds in this
case, as we will see from the mechanism of construction that the duality holds separately in each
twisted sector; and within a twisted sector, $\cN=2$ supersymmetry  is preserved.

\bigskip

The generic form of this partition function is quite lengthy but useful. We note, as an index of
the various blocks, the corresponding degrees of freedom. Noting for brevity $h_3=-h_1-h_2$, it
reads :

\ba \non Z_{\cN=1} & = & \int_F
\frac{d^2\tau}{\tau_2^2}~\frac{\tau_2^{-1}}{\eta^{12}\bar\eta^{24}}~ \frac 1{2^2} \sum_{h_i,g_i}
\left(\frac 12 \sum_{a,b} (-)^{a+b+ab}
\Th{a}{b}\Th{a+h_1}{b+g_1}\Th{a+h_2}{b+g_2}\Th{a+h_3}{b+g_3}\right)_{\psi^\mu, \chi}\\
\label{partition_function} &\times & \left(\frac 12 \sum_{\epsilon,\xi} \Thbar{\epsilon}{\xi}^5
\Thbar{\epsilon+h_1}{\xi+g_1} \Thbar{\epsilon+h_2}{\xi+g_2}
\Thbar{\epsilon+h_3}{\xi+g_3}\right)_{\bar\psi^{1\dots 5},\bar\eta^{1,2,3}}\\
\non & \times & \left(\frac 12 \sum_{H_1,G_1}\frac 12 \sum_{H_2,G_2} (-)^{H_1G_1 + H_2G_2}
\Thbar{\epsilon+H_1}{\xi+G_1}^4
\Thbar{\epsilon+H_2}{\xi+G_2}^4\right)_{\bar\phi^{1\dots 8}}\\
\non & \times & \left(\sum_{s_i,t_i} \Gamma_{6,6}\left[^{h_i | s_i}_{g_i |
t_i}\right]\right)_{(y\omega \bar y\bar \omega)^{1\dots 6} } \times e^{i\pi
\Phi(\gamma,\delta,s_i,t_i,\epsilon,\xi,h_i,g_i,H_1,G_1,H_2,G_2)}\ea where the internal
twisted/shifted $(6,6)$ lattice is given by \ba \non \Gamma_{6,6}\left[^{h_i | s_i}_{g_i |
t_i}\right] & = & \frac 1{2^6} \sum_{\gamma_i,\delta_i}\,
\left(\left|\Th{\gamma_1+h_1}{\delta_1+g_1}\right|\,\left|\Th{\gamma_1}{\delta_1}\right|
\,(-)^{\gamma_1 t_1+\delta_1 s_1 + s_1 t_1}\right)_{(y\omega \bar y\bar \omega)^1}\\
 & & \times \left(\left|\Th{\gamma_2+h_1}{\delta_2+g_1}\right|\,\left|\Th{\gamma_2}{\delta_2}\right|
\,(-)^{\gamma_2 t_2+\delta_2 s_2 + s_2 t_2}\right)_{(y\omega \bar y\bar \omega)^2}\\
\non & & \times \left(\left|\Th{\gamma_3+h_2}{\delta_3+g_2}\right| \,
\left|\Th{\gamma_3}{\delta_3}\right|
\,(-)^{\gamma_3 t_3+\delta_3 s_3 + s_3 t_3}\right)_{(y\omega \bar y\bar \omega)^3}\\
\non & & \times \left(\left|\Th{\gamma_4+h_2}{\delta_4+g_2}\right| \,
\left|\Th{\gamma_4}{\delta_4}\right|
\,(-)^{\gamma_4 t_4+\delta_4 s_4 + s_4 t_4}\right)_{(y\omega \bar y\bar \omega)^4}\\
\non & & \times \left(\left|\Th{\gamma_5+h_3}{\delta_5+g_3}\right| \,
\left|\Th{\gamma_5}{\delta_5}\right|
\,(-)^{\gamma_5 t_5+\delta_5 s_5 + s_5 t_5}\right)_{(y\omega \bar y\bar \omega)^5}\\
\non & & \times \left(\left|\Th{\gamma_6+h_3}{\delta_6+g_3}\right| \,
\left|\Th{\gamma_6}{\delta_6}\right| \,(-)^{\gamma_6 t_6+\delta_6 s_6 + s_6 t_6}\right)_{(y\omega
\bar y\bar \omega)^6}.\ea Here $e^{i\pi\Phi}$ is a global phase whose effect is to implement the
various GGSO projections acting on the spectrum of this theory. Following  the formalism of
\cite{FF}, these GGSO projections are equivalently defined by the coefficients $C_{(v_i|v_j)}
\equiv [v_i|v_j]$,
where $v_i$ and $v_j$ are the vectors of \eqref{full_basis}.\\
This phase is required to satisfy modular invariance constraints, that is, it must be invariant
under the following transformations :

\be \label{modular_tranformations} \tau \to \tau+1  \Rightarrow \left\{\begin{array}{lcr} (a,b) &
\to & (a,a+b+1)\\
(\gamma_i,\delta_i) & \to & (\gamma_i,\gamma_i+\delta_i+1)\\
(\epsilon,\xi) & \to & (\epsilon, \epsilon+\xi+1)\\
(h_i,g_i) & \to & (h_i,h_i+g_i)\\
(H_i,G_i) & \to & (H_i,H_i+G_i)\\
(s_i,t_i) & \to & (s_i,s_i+t_i)\\
\end{array} \right. ~~;~~
\tau \to -1/\tau  \Rightarrow \left\{\begin{array}{lcr} (a,b) & \to & (b,a)\\
(\gamma_i,\delta_i) & \to & (\delta_i,\gamma_i)\\
(\epsilon,\xi) & \to & (\xi,\epsilon)\\
(h_i,g_i) & \to & (g_i,h_i)\\
(H_i,G_i) & \to & (G_i,H_i)\\
(s_i,t_i) & \to & (t_i,s_i)\\
\end{array} \right.~~. \ee

\bigskip
Here we may make some remarks.

\bi \item The global phase $\Phi$ does not depend on the spin structure of the space-time fermions,
$(a,b)$. This is necessary to preserve $\cN=1$ supersymmetry; otherwise supersymmetry is
spontaneously broken, as the gravitini acquire a mass. We will not consider this mechanism here.
Note however that the construction of a realistic model also requires such a breaking.

\item We want to emphasize the physical meaning of the parameter $\epsilon$ in the expression
\eqref{partition_function}. As the $\bar\psi$ block corresponds to the representations of $SO(10)$,
$\epsilon$ is the associated chirality~: spinorials of $SO(10)$ have $\epsilon=1$, whereas
vectorials have $\epsilon=0$. We will relate this later to the right-moving SCFT of the model;
breaking this SCFT will be done by assuming a non-trivial dependence of the global phase $\Phi$ of
the spin-structure $(\epsilon, \xi)$.

\item The inclusion of $(s_i,t_i)$ performs additional shifts on the six (fermionized)
internal dimensions compactified on $T^6/\Z_2\times \Z_2$. These shifts correspond to the presence
of the sets $(e_i)$ in the parity basis; similarly, the twisting parameters $(H_i,G_i)$ account for
the presence of the sets $z_i$. Coupling these parameters to various spin structures by a suitable
form of the phase $\Phi$ will generate the Scherk-Schwarz symmetry breakings we will consider.

\ei

\subsection{$SO(10)$ models as Gepner-map duals of Type II models}

The model we have considered above is in fact obtained directly from a Type II model by a map
introduced in \cite{GepnerLust}. This map defines a correspondence between a heterotic model and a
Type II model by the following construction.

\bigskip

\noindent If we label $B_{\lambda=1,2,3,4}$ the four characters of $SO(8)$ $O_8,V_8,S_8,C_8$, one
can write a generic Type II partition function in the following form

\be Z_{II} = \frac{1}{\tau_2^4 \eta^8 \bar\eta^8} \sum_{\lambda,\bar\lambda}B_\lambda \bar
B_{\bar\lambda} Z_{\lambda,\bar\lambda} \ee Here, $Z_{\lambda,\bar\lambda}$ account for the
spin-statistics of the model and, in the case of compactified theories, for the internal lattices.
The general procedure\footnote{There exists a second solution, which is the replacement by $SO(32)$
characters.}  is then to replace the $SO(2d)$ characters of the right-moving side of the theory by
$SO(8+2d) \times E_8$ characters, so that the modular properties of the partition function are
preserved. The product only involves the singlet character of $E_8$, whereas the map for the
$SO(2d)$ characters is done as follows : \be \bar O_{2d} \to \bar V_{2d+8},~~\bar V_{2d} \to \bar
O_{2d+8},~~\bar S_{2d} \to -\bar S_{2d+8},~~\bar C_{2d} \to -\bar C_{2d+8}. \ee In particular, for
the usual IIA and IIB space-time fermions blocks, $d=4$ and the replacement is done by

\ba \frac 12 \sum_{\ab,\bb} (-)^{\ab+\bb} \Thbar \ab\bb^4 & \to & \left[\frac 12 \sum_{\ab,\bb}
(-)^{\ab\bb}\Thbar \ab\bb^8\right]\times \frac 12 \sum_{\gb,\db}\Thbar \gb\db^8\\
 \frac 12 \sum_{\ab,\bb} (-)^{\ab+\bb+\ab\bb} \Thbar \ab\bb^4 & \to & \left[\frac 12 \sum_{\ab,\bb}
\Thbar \ab\bb^8\right]\times \frac 12 \sum_{\gb,\db}\Thbar \gb\db^8  \ea We see that the reversal
of the sign of the fermionic characters breaks the usual spin-statistics, so that, from a
space-time point of view, this operation has traded a supersymmetric sector for a purely bosonic
sector. Following our notations for the free fermionic degrees of freedom and their obvious
extension to Type II models, the mapping Type II $\to$ Heterotic is done by replacing the free
fermions of Type II $\{\bar\psi^\mu,\bar\chi^{1\dots 6}\}$ by the free fermions of the heterotic
$\{\bar \psi^{1\dots 5},\bar\eta^{1,2,3},\bar\phi^{1\dots 8}\}$. Also note that in both Type IIA
and Type IIB cases, the obtained block is in fact a second copy of the singlet of $E_8$,
which signals an enhancement of $SO(16)$ to $E_8$.\\
Carrying out the $\Z_2 \times \Z_2$ orbifold on both of these models, we see that the heterotic
model we consider in this paper is no other than the Gepner-map of a Type II
$\cN_4=2$ model, \emph{via} the mapping\\

\vspace{2cm}

\be \frac 12 \sum_{\ab,\bb} (-)^{\ab+\bb+\ab\bb}\Thbar {\ab}{\bb}\Thbar {\ab+h_1}{\bb+g_1}\Thbar
{\ab+h_2}{\bb+g_2}\Thbar {\ab+h_3}{\bb+g_3} \longrightarrow
~~~~~~~~~~~~~~~~~~~~~~~~~~~~~~~~~~~~~~~\ee

$$~~~~~~~~~~~~~~~~~~~~~~~~~~~~~~~~~~~~~\left[\frac 12 \sum_{\ab,\bb} \Th \ab\bb^5
\Thbar {\ab+h_1}{\bb+g_1}\Thbar {\ab+h_2}{\bb+g_2}\Thbar {\ab+h_3}{\bb+g_3}\right]\times \frac 12
\sum_{\gb,\db}\Thbar \gb\db^8.$$ One recognizes the block of \eqref{partition_function}
corresponding to the $\bar\psi$'s and $\bar\eta$'s. The second block accounts for an $E_8$ gauge
group formed by the complex fermions $\bar\phi^{1\dots 8}$; generically, this group will be broken
due to the inclusion of the sets $z_1$ and $z_2$ in our construction.

Out of the two four-dimensional supersymmetries of the Type II model, only the left-moving one is
still present in the heterotic; however, the right-moving superconformal algebra survives the
mapping. This is nothing but the embedding of the spin connection of Type II models into the
connection of the corresponding heterotic ones. Then, this superconformal algebra does not give
birth to a space-time SUSY, but relates spinors to vectors, belonging to representations which are
now of the internal $SO(10)$ spanned by the $\bar\psi$'s. The survival of this symmetry will
guarantee the existence at the massless level of what were formerly right-moving gravitinos and are
now gauge bosons in a spinorial of $SO(10)$ : then, $SO(10)\times U(1)^3$ gets enhanced to
$E_6\times U(1)^2$. This enhancement comes as no surprise from the Calabi-Yau point of view : the
general embedding of spin-connection into gauge connection singles out a subalgebra $SU(3)$ inside
the first $E_8$, corresponding to the holonomy of the compactification manifold. The anomaly
cancellation mechanism \cite{GreenSchwarz} then requires that we switch on background values for
this $SU(3)$, and the surviving gauge group is $E_6$, coming from the embedding $SU(3) \times E_6
\subset E_8$. Of course, the Cartans of $SU(3)$ still define a gauge group $U(1)^2$, so that, in
the presence of a right-moving $N=2$ SCFT, we indeed find a gauge group $E_6 \times U(1)^2 \times
E_8$. This is realized explicitly in our constructions.

\bigskip

Note that this procedure underlines the naturalness of the appearance of a gauge group $SO(10)$ in
$\cN=1$ realistic theories : the Type II right-moving fermionic block made out of $S,V,C$
representations of the Lorentz group $SO(8)$ is traded for a block made out of $E_8$ characters.
The $\Z_2\times \Z_2$ orbifold required to break the four-dimensional supersymmetry $\cN=4 \to 1$
is forced by consistency to act on this $E_8$, generically breaking it to $E_6 \times U(1)^2$.

\bigskip

We will now enumerate the sectors from which we will be able to build massless states, and identify
their interpretation as twisted sectors of the $\cN=4 \to \cN=1$ $\Z_2\times \Z_2$ orbifold.

\section{Spectrum of the model; superconformal $x$-map and its spontaneous breaking}

\subsection{$\Z_2\times \Z_2$ twisted sectors}\label{twisted_sectors}

It is pretty straightforward to check that the $\cN=1$ supersymmetric partner of a state built on
some vacuum $|\alpha\ket$ will come from the vacuum $|\alpha+S\ket$. Here, we will therefore
restrain our enumeration to the bosonic vacua. Apart form the pure NS vacuum, states can be built
from the following sets :
\begin{itemize}
 \ittem the 16 twisted sectors
$\left|\cB_{\lambda_3\lambda_4\lambda_5\lambda_6}^1\right\ket=\left|b_1+ \sum_{i=3}^6 \lambda_i
e_i\right\ket$, where $\lambda_i=0\ {\rm or}\ 1$;
 \ittem the 16 twisted sectors
$\left|\cB_{\lambda_1\lambda_2\lambda_5\lambda_6}^2\right\ket=\left|b_2+ \sum_{i=1,2,5,6} \lambda_i
e_i\right\ket$, where $\lambda_i=0\ {\rm or}\ 1$;
 \ittem the 16 twisted sectors
$\left|\cB_{\lambda_1\lambda_2\lambda_3\lambda_4}^3\right\ket=\left|b_3+ \sum_{i=1}^4 \lambda_i
e_i\right\ket$, where $\lambda_i=0\ {\rm or}\ 1$;

 \ittem the sectors $\left|\alpha+x\right\ket$, where $\alpha$ is any of the sectors described
above;
 \ittem the sectors $\left|z_1\right\ket$, $\left|z_2\right\ket$, $\left|z_1+z_2\right\ket$.

\end{itemize}

\noindent To properly distinguish a particle from its anti-particle, it will be handy to consider
instead the fermionic sectors $B\equiv S+\cB$, so as the space-time chirality appears in a clear
way. We will then restrain ourselves to considering positive $\psi^\mu$-helicity states. In the
following, we will denote $|B^1\ket$ (and similarly for $|B^2\ket$, $|B^3\ket$) a generic sector
$\left|B_{\lambda_1\lambda_2\lambda_3\lambda_4}^1\right\ket$, and more generally $|B\ket$ an
arbitrary twisted sector. The $|B\ket$ sectors are in one-to-one correspondence with
the fixed points of the $\Z_2\times \Z_2$ orbifold transformation.\\
Let us make some comments :

\bi

\item In the following, we will pay no attention to the sectors $|z_1\ket$, $|z_2\ket$,
$|z_1+z_2\ket$, which can lead to additional gauge bosons. The minimal gauge group  is $SO(8)
\times SO(8)$; as pointed out in \cite{FKR}, appropriate choice of the GGSO phases ensures that
this gauge group is not enhanced, and that no mixed\footnote{By mixed states, we mean states
charged under both the ``observable''  $SO(10)$ or $E_6$ and the ``hidden'' gauge group containing
the $SO(8)\times SO(8)$.} massless states appear. In the following, we will assume these
no-enhancement hypotheses, which state that there exists $e_i$ and $e_j$, $i\neq j$, such as
$[e_i|z_1]=-1$ and $[e_j|z_2]=-1$. This choice projects out any would-be gauge bosons that would
enhance $SO(8)\times SO(8) \to SO(16)$; the largest enhancement one can have in that case is a
$SO(8) \to SO(9)$, which can also be eliminated by allowing one more $i$ such as $[e_i|z_1]=-1$; at
any rate, there is no mixing between the ``observable'' gauge and the ``hidden'' gauge.

\item The spinor-vector duality finds its root from the fact that if $|\alpha\ket$ is a
relevant vacuum to build massless states, so is $|\alpha+x\ket$. This correspondence is the
superconformal ``$x$-map'' $|B\ket \mapsto |B+x\ket$ pointed out in \cite{xmap}. It is obvious that
if (the excitations of) $|\alpha\ket$ are in the vectorial of the $SO(10)$ induced by the 5 complex
fermions $\bar\psi^{1\dots 5}$, then $|\alpha+x\ket$ will belong to a spinorial of the same group;
the $x$-map being an involution, the converse is also true. What is at stake is then to find, given
a set of GGSO projections, which sectors will survive; and for each theory, describe the dual
theory in terms of the effects of its various GGSO projections.

\item An important case of figure brings a self-dual case. When preserving the $N=(0,2)$
superconformal field theory, the $SO(10)_{\bar\psi} \times U(1)_{\bar\eta}$, where
$U(1)_{\bar\eta}$ is the diagonal $U(1)$ induced by $\bar\eta^{1,2,3}$, is lifted to $E_6$. In this
case, the vectorial {\bf 10} and the spinorial {\bf 16} of $SO(10)$ (resp the anti-spinorial
$\overline{{\bf 16}}$) are grouped in the fundamental {\bf 27} (resp. $\overline{{\bf 27}}$) of
$E_6$, which decomposes as ${\bf 27} \to {\bf 10}\oplus {\bf 16} \oplus {\bf 1}$ (resp.
$\overline{{\bf 27}} \to {\bf 10}\oplus \overline{{\bf 16}} \oplus {\bf 1}$).


\ei
\subsection{The $x$-map and superconformal algebra in representations of $SO(10)$}

To begin with, we will restrain ourselves to consider only one twisted sector, namely
$B^1_{0000}=S+b_1$. We will note the associated ground state $|B^1_{0000}\ket$. Our results will
easily be extended to any of the 48 twisted sectors detailed above. The untwisted sector, built out
of the pure Neveu-Schwarz ground state, gives the gauge bosons of the gauge group, but not the
spinorial/vectorial representations we are
interested in.\\
The $B^1_{0000}$ vacuum is then written as

\be B^1_{0000}~:~\Sp\left(\psi^\mu,\chi^{1,2},y^{3\dots 6}\right)\otimes \Sp\left(\bar y^{3\dots
6},\bar \psi^{1\dots 5},\bar \eta^1\right) \ee and the addition of the sector $x$ brings the vacuum

\be B^1_{0000}+x~:~\Sp\left(\psi^\mu,\chi^{1,2},y^{3\dots 6}\right)\otimes \Sp\left(\bar y^{3\dots
6},\bar \eta^{2,3}\right). \ee Here, one may make a few remarks, which will be valid for any of the
48 twisted sectors. Firstly, due to the presence of 8 left-moving and 16 right-moving real fermions
obeying Ramond boundary conditions, the sector $|B^1_{0000}\ket$ is massless by itself, and
contains spin-fields made out of the $SO(10)$ fermions $\bar\psi$; it therefore induces a spinorial
of $SO(10)$. On the other hand, the sector $|B^1_{0000}+x\ket$ has 8 left-moving and 8 right-moving
Ramond real fermions, so that its ground energies read
$$M_L^2 = 0;~~~~M_R^2 = -\frac 12.$$
A massless state will then be reached when exciting this ground state by a weight $1/2$
right-moving fermionic oscillator. If we wish to consider states charged under $SO(10)$, this
excitation has to be taken to be $\bar\psi^i_{-1/2}$, and the resulting state lies in a vectorial
representation of $SO(10)$. Therefore, the $x$-map links vectorials to spinorials of $SO(10)$.
Obviously, the $x$-map arises as the right-moving part of the $N=(2,2)$ superconformal field theory
that is still present after the Type II $\to$ Heterotic Gepner-map, and acts inside the gauge
group, due to the embedding of the spin connection into the gauge connection.

\bigskip

As in the case of spontaneous breaking of supersymmetry, a spontaneous breaking of the $x$-map will
amount to projecting out from the spectrum spinorial or vectorial representations of $SO(10)$,
giving different masses to the two partners. In terms of the free fermionic construction, this
situation is reflected in the fact that states from the massless sector $|B\ket$ (resp. $|B+x\ket$)
will be projected out, whereas states from the sectors $|B+e_i\ket$ (resp. $|B+x+e_i\ket$) will be
preserved. These sectors are massive and are naturally interpreted as the twisted sector of the
freely-acting orbifold based on the half-shift of the coordinate $X^i$. We see that the net effect
of this action is that the sectors $|B\ket$ (resp. $|B+x\ket$) will get a mass, whereas the sectors
$|B+x\ket$ (resp. $|B\ket$) will remain massless. We carry out an explicit example of such a mass
lift in the next subsection; as one can expect, it crucially relies on a careful choice of the GGSO
projections.

\subsection{Implementing the $e_i$-generated freely-acting orbifolds}

In this subsection, we briefly recall some useful results about twisted/shifted lattices. The usual
equivalence between a compact boson taken at the fermionic point and two left-moving plus two
right-moving real fermions is easily extended to orbifold partition
functions of each theory.\\
When we consider two internal dimensions , the $\vt$-function form of a zero-mode lattice
$\Gamma_{2,2}$, taken at the enhanced symmetry (or fermionic) point (denoted $f.p.$)

\be \Gamma_{2,2}\Big|_{f.p.} = \left( \frac 12 \sum_{\gamma,\delta}
\left|\Th{\gamma}{\delta}\right|^2\right)^2 \ee is generalized to the orbifold version of the
theory. When one implements the non-freely-acting $\Z_2$ orbifold $X^{1,2} \to -X^{1,2}$, whose
twisting parameters will be denoted $(h,g)$, as well as the two freely-acting $\Z_2$ orbifolds
$X^{1,2} \to X^{1,2}+\pi$, whose shifting parameters will be noted $(s_1,t_1,s_2,t_2)$, the lattice
sum is modified as

\ba \label{twist_shift_22} \Gamma_{2,2}\left[^{h|s_1,s_2}_{g|t_1,t_2}\right] \Big|_{f.p.}& = &
\frac 14\sum_{\gamma_{1,2},\delta_{1,2}} (-)^{\gamma_1 t_1 + \delta_1 s_1 + s_1t_1}(-)^{\gamma_2
t_2 +
\delta_2 s_2 + s_2t_2}\\
 \non & \times & \left|\Th{\gamma_1+h}{\delta_1+g} \Th{\gamma_1}{\delta_1}
\Th{\gamma_2+h}{\delta_2+g} \Th{\gamma_2}{\delta_2}\right| \ea Therefore, implementing in the above
partition function the freely-acting orbifolds (in this case, half-way shifts) corresponding to the
sets $e_i$ only amounts to inserting the phases $(-)^{\gamma t + \delta s + st}$. For now, we have
just shifted the internal $\Gamma_{6,6}$ lattice, independently of the rest of the spectrum. The
corresponding orbifold is the $\Z_2$-translation along each circle of the internal space.

\bigskip

If we wish to couple this shift to other states of the theory, we must introduce a phase relating
the shift parameters $(s_i,t_i)$ to the spin structures of the states we want to act on. Such a
freely-acting orbifold takes the form $(-)^Q \cdot T^i$, where $T^i$ is the $\Z_2$-translation of
the $i^{{\rm th}}$ coordinate $X^i \mapsto X^i + \pi R^i$, and $(-)^Q$ is the parity operator
associated to the spin structure we are considering (generalizing the usual fermion counting
operator $(-)^F$, which would correspond to coupling to the spin-structure of the space-time
fermion spin structure $(a,b)$).

One can carry out the calculation of the partition function corresponding to this orbifold, by
inserting the projection operator in the computation of the trace over physical states and adding
the contribution of the twisted sector. The result is that this orbifold is done by simply adding a
cocycle in the partition function. As an example, if we consider a $\Gamma_{1,1}$ lattice coupled
to some spin structure $(\epsilon,\xi)$, the modification is made as follows :

\be \non \label{orbifold-form} Z = [...] \frac R{\sqrt{\tau_2}}\sum_{\tilde m,n}
\exp\left[-\frac{\pi R^2}{\tau_2}|\tilde m+n\tau|^2\right] \ee

\ba & \longrightarrow & [...]\times\frac 12 \sum_{h,g} (-)^{\epsilon g + \xi h + gh} \frac
R{\sqrt{\tau_2}} \sum_{\tilde m,n} \exp\left[-\frac{\pi R^2}{\tau_2}\left|\left(\tilde
m+\frac g2\right)+\left(n+\frac h2\right)\tau\right|^2\right]\\
\non & = & [...]\times  \sum_{h,g} (-)^{\epsilon g + \xi h + gh}\, \Gamma_{1,1}[^h_g]\left(\frac
 R2\right)\ea
where $\Gamma_{1,1}[^h_g]$ is the shifted $\Gamma_{1,1}$ lattice

\be \Gamma_{1,1}[^h_g] = \frac R{\sqrt{\tau_2}} \sum_{\tilde m,n} \exp\left[-\frac{\pi
R^2}{\tau_2}\left|\left(2\tilde m+ g\right)+\left(2n+h\right)\tau\right|^2\right] \ee and the
overall $[...]$ refers to all the other blocks of the partition function, which are unchanged in
the process.\\ Setting $R_{SS}=R/2$, we recover the well-known fact that this mechanism is
equivalent to performing a stringy Scherk-Schwarz compactification, which is done by coupling the
internal dimension to the $SO(10)$ helicity current \cite{StringySS} $$\oint
(\bar\psi^1)^\dag\,\bar \psi^1.$$ Such a task is achieved by inserting in the concerned partition
function block the cocycle

\be \label{SS-form} (-)^{\epsilon\tilde m + \xi n + \tilde m n},\ee where now $\tilde m $ and $n$
are the momentum/winding numbers of the string state along the radius $R_{SS}$ \cite{StringySS}.
Looking at the expressions \eqref{twist_shift_22} and \eqref{orbifold-form}, one sees that, since
the internal shift parameters of the internal dimensions are no other than $(s_i,t_i)$ that the
coupling of the internal shifted lattice to the $SO(10)$ spin-structure $(\epsilon,\xi)$ will be
done by inserting a phase of the form

\be (-)^{\epsilon t_i + \xi s_i + s_it_i}. \ee It is worth noting that this coupling indeed lifts
the mass of the states according to their chirality $\epsilon$ : by considering the insertion of
the Scherk-Schwarz cocycle \eqref{SS-form}, a Poisson resummation of the modified lattice

\be \frac{R_{SS}}{\sqrt{\tau_2}} \sum_{\tilde m,n} (-)^{\epsilon \tilde m + \xi n + \tilde m n}
\exp\left[-\frac{\pi R_{SS}^2}{\tau_2}|\tilde m+n\tau|^2\right] \ee shows that the string states
now have momentum and winding numbers \be \left(m-\frac \epsilon 2-\frac n2,n\right)\ee which
signals a mass lifting in the $\epsilon=1$ sector. This procedure is of course encoded in the basic
form of the fermionic construction and does not require further elaboration : it is related to the
values of the discrete torsions $[e_i|B]$ and $[e_i|B+x]$, where $B$ is an arbitrary twisted sector
of the theory.

\subsection{Breaking the $x$-symmetry with the freely-acting orbifold $e_i$}

We start by considering the two sectors already written above, which read, in terms of spin-fields

\be B^1_{0000}~:~\Sp\left[(\psi^\mu)_{+},(\chi^{12})_{\epsilon_2},(y^{34})_{\epsilon_3},
(y^{56})_{\epsilon_4}\right]\otimes \Sp\left[(\bar y^{34})_{\bar \epsilon_1} ,(\bar y^{56})_{\bar
\epsilon_2},(\bar \psi^{1\dots 5})_{\bar \epsilon_3},(\bar \eta^1)_{\bar \epsilon_4}\right] \ee

\be B^1_{0000}+x~:~\Sp\left[(\psi^\mu)_{+},(\chi^{12})_{\epsilon_2},(y^{34})_{\sigma_3},(y^{56})_{
\sigma_4}\right]\otimes \Sp \left[(\bar y^{34})_{\bar \sigma_1}, (\bar y^{56})_{\bar
\sigma_2},(\bar \eta^2)_{\bar \sigma_3} ,(\bar \eta^3)_{\bar \sigma_4}\right] \ee where the
$\epsilon_i$, $\bar\epsilon_i$, $\sigma_i$, $\bar\sigma_i$ are the helicities
of the spin-fields.\\
As discussed above, the physical states of the sector $B^1_{0000}+x$ we are interested in are
obtained by exciting the vacuum with a weight $1/2$ $\bar\psi$ oscillator :

\be \Sp\left[(\psi^\mu)_{+},(\chi^{12})_{\epsilon_2},(y^{34})_{\sigma_3},(y^{56})_{\sigma_4}\right]
\otimes \Big[\bar\psi^i_{-1/2}\Big]\,\Sp \left[(\bar y^{34})_{\bar \sigma_1}, (\bar y^{56})_{\bar
\sigma_2},(\bar \eta^2)_{\bar \sigma_3} ,(\bar \eta^3)_{\bar \sigma_4}\right] \ee The relevant GGSO
projections to carry out in this example as those arising from the sets $S,\ S+b_1,\ b_2,\
(e_i)_{i=1\dots 6}$. The $F$-projection is redundant with the $S+b_1$-one. The $z_i$-projections do
not change the features of the spectrum in the sector $B^1_{0000}$ as soon as we assume that they
do not project the whole sector out; we
will, for now, neglect them.\\
Equivalently, we will find it handy to consider instead, on a sector
$B_{\lambda_3\lambda_4\lambda_5\lambda_6}^1$  the projections induced by the sets

\be S,~~~~~~~~S+b_1,~~~~~~~~\tilde b_2=S+b_2+(1-\lambda_5)e_5+
(1-\lambda_6)e_6,~~~~~~~~(e_i)_{i=1\dots 6}~. \ee Recall that, as
$|B^1_{\lambda_3\lambda_4\lambda_5\lambda_6}\ket$ are fermionic sectors, the constraints to be met
are $(-)^\alpha = -(\alpha|B^1_{\lambda_3\lambda_4\lambda_5\lambda_6})$, where $\alpha$ is one of
the sets
above.\\

Initially, the sectors $B^1_{\lambda_3\lambda_4\lambda_5\lambda_6}$ have $2^{12}$ degrees of
freedom. Carrying out the $S$, $S+b_1$, $\tilde b_2$, $(e_{3\dots 6})$ projections cut the number
of physical states down to $2^5=32$. Noticing that

\be \label{b2tilde} \tilde b_2 \cap B^1_{\lambda_3\lambda_4\lambda_5\lambda_6} =
\{\psi^\mu|\bar\psi^{1\dots 5}\} ,\ee we see that, as the $\psi^\mu$ helicity has been fixed, this
GGSO projection implies that the spectrum of states inside the sectors
$B^1_{\lambda_3\lambda_4\lambda_5\lambda_6}$ is chiral with respect to the group $SO(10)$. Such a
feature crucially depends on the presence of the set $b_2$ in our construction; this is consistent
with the fact that the presence of a chiral matter spectrum requires $\cN=1$ space-time
supersymmetry.

Now we look at the effect of the $e_1$ and $e_2$ projections, first restricting our attention to
$B^1_{0000}$. The latter survives the $e_1$ projection if $[e_1|B^1_{0000}]=-1$; otherwise the
entire sector $|B^1_{0000}\ket$ is projected out. However, in the latter case, as mentioned
earlier, one has to consider the massive sector $|B^1_{0000}+e_1\ket$. The spin field accounting
for this Ramond ground now has an initial degeneracy of $2^{14}$; carrying out the $S$, $S+b_1$,
$\tilde b_2$, $(e_{1,3\dots 6})$ projections cut the number of degrees of freedom to $2^6$. This
time, the various projections are not able to fix the $SO(10)$-chirality of the massive state,
since

\be  \tilde b_2 \cap \left(B^1_{\lambda_3\lambda_4\lambda_5\lambda_6} +e_1\right)=
\{\psi^\mu,\omega^1|\bar\omega^1,\bar\psi^{1\dots 5}\} .\ee This is consistent with the fact that
when fixing the space-time spin, we still have a degeneracy in the representations ${\bf 16}$ and
$\overline{{\bf 16}}$ of $SO(10)$, which is mandatory for these representations to be massive.

\bigskip

\noindent The superconformal partner of $|B^1_{0000}\ket$ is $|B^1_{0000}+x\ket$; this sector
contains vectorial representations of $SO(10)$. Let us recall that, from the usual constraints of
the free fermionic models, the discrete torsion coefficients we are interested in obey, for $i=1,2$
:

\be \label{GGSO_product} [B^1+x|e_i]=[B^1|e_i][x|e_i].\ee Therefore, if we set $[x|e_i]=1$, the
sector $|B^1_{0000}+x\ket$ will behave in the same way as $|B^1_{0000}\ket$ with respect to the
$e_i$ projections. If $[B^1_{0000}|e_i]=1$, the twisted sector will be projected out as a whole,
regardless of the spinorial/vectorial character of the representations; if $[B^1_{0000}|e_i]=-1$,
both spinors and vectors will survive.

\bigskip

Up to now, we have thus not been able to discriminate between spinorial and vectorial
representations of $SO(10)$ lying in the same twisted sector. As one can expect, this will be done
by acting on the value of the discrete torsion $[x|e_i]$. Indeed, let us again place ourselves in
the twisted sector $|B^1_{0000}\ket$, and its vectorial counterpart $|B^1_{0000}+x\ket$. The same
reasoning as before, and the use of the equation \eqref{GGSO_product}, yields the following rules
of survival (we recall that $\delta_B=-1$ for any fermionic twisted sector):

\bi
 \item when $[B^1_{0000}|e_i]=-1$ and $[x|e_i]=1$, both sectors $|B^1_{0000}\ket$ and
$|B^1_{0000}+x\ket$ survive at the massless level;
 \item when $[B^1_{0000}|e_i]=1$ and $[x|e_i]=1$, both sectors $|B^1_{0000}\ket$ and
$|B^1_{0000}+x\ket$ are projected out;
 \item when $[B^1_{0000}|e_i]=-1$ and $[x|e_i]=-1$, $|B^1_{0000}\ket$ survives and
$|B^1_{0000}+x\ket$ is projected out;
 \item when $[B^1_{0000}|e_i]=1$ and $[x|e_i]=-1$, $|B^1_{0000}\ket$ is projected out and
$|B^1_{0000}+x\ket$ survives. \ei Now that we know how to manipulate each twisted sector, we can
start to explore the duality. Note that the list of ingredients at our disposal is quite simple and
handy.

\bigskip

We are dealing with three twisted planes, in which four left-moving and four right-moving fermions
picked among the fermionized coordinates $(y^i\omega^i)(\bar y^i\bar\omega^i)$ are in Ramond
boundary conditions. These fermions carry indices $(i_1,i_2,i_3,i_4)=(3,4,5,6)$ for the $B^1$
family, $(1,2,5,6)$ for the $B^2$ family, and $(1,2,3,4)$ for the $B^3$ family. We can act on these
twisted sectors by making the freely-acting orbifold generated by the set $e_i$ act in a
non-trivial way on them. Then one sees that, to be able to project out states, one must consider
the action of the sets $e_i$ and $e_j$, where $i$ and $j$ are different from $i_{1\dots 4}$;
otherwise, the $e_i$-projection's effect is to choose the internal chiralities of the corresponding
spin-field. Moreover, if $i$ is one of the four indices $i_{1\dots 4}$, the sector
$B+e_i$ is not massive, but rather another twisted sector of the same plane.\\
Then two projections have to be considered for each twisted plane. In the following, we will be
interested in the $B^1$ plane, so that we will consider the orbifolds induced by $e_1$ and $e_2$.
This fact is not surprising : in the $B^1$ plane, the physics is independent of the volume of the
four internal coordinates corresponding to the fermions $(y\omega\vert\bar y\bar\omega)^{3456}$;
therefore, a spontaneous breaking of symmetry in this plane must be constructed out of the two last
internal coordinates, as the value of the mass gap will depend on the size of these coordinates. Of
course, in this paper we will encounter no such dependence, as all moduli are set at the fermionic
point; however, a deformation of these models would make this feature clear.

Finally, to compute the action of the orbifolds $e_1$ and $e_2$ on one arbitrary sector of the
first twisted plane $|B^1_{\lambda_3\lambda_4\lambda_5\lambda_6}\ket$, we remark that the usual
constraints of the fermionic construction impose

\be \left[B^1_{\lambda_3\lambda_4\lambda_5\lambda_6}|e_i\right] = [b_1+S|e_i]\prod_{j=3}^6
[e_j|e_i]^{\lambda_j}~~~~~~~~~i=1,2.\ee Knowing all the coefficients $[e_i|e_j]$ , which are part
of the definition of the model, we are then able to repeat the above reasoning to deduce the action
of $e_1$ and $e_2$ projections on $|B^1_{\lambda_3\lambda_4\lambda_5\lambda_6}\ket$ and
$|B^1_{\lambda_3\lambda_4\lambda_5\lambda_6}+x\ket$.

\subsection{The $z_i$ projections}

The case of the $z_i$ projections is in many ways similar to the case of the $e_i$'s. This time, as
we have, for any twisted sector $B$ of the theory $B \cap z_1 = B\cap z_2 = \varnothing$, any
non-trivial discrete torsion turned on for the $z$ sets will have an effect on the three twisted
planes. One can derive all the rules in a similar way as for the $e_i$'s : the $z_i$ projections
can be taken to break the $x$ superconformal CFT or not, and various combinations of hypotheses on
the GGSO yields various cuts in the spectrum of the theory. As this case is identical to the
$e_{1,2}$ orbifolds, the rules of the previous subsection apply.

\bigskip

\noindent We will often omit the  $z_i$ projections, to which most of the rules we derive for the
$(e_i)$ projections similarly apply. We will actually specifically need them to perform further
cuts in the spectrum, giving us the possibility to restrain the number of representations present
in our models.

\section{Construction of dual pairs of models} \label{construction}

\subsection{A class of self-dual models : the $E_6$ models}

As we mentioned previously, since in $E_6$ models the spectrum arranges itself in fundamental
representations ${\bf 27}$ and $\overline{{\bf 27}}$, these models are trivially self-dual.

The gauge group $E_6$ is present in a model if and only if the $x$-map is unbroken. This is
equivalent to requiring that the freely-acting orbifolds do not break the right-moving part of the
$N=(2,2)$ superconformal algebra of the initial model. In terms of discrete torsion coefficients,
this condition is encoded in the equality

\be \forall ~ i=1\dots 6,~~~~ [x|e_i]=1;~~~~~ [x|z_{1,2}]=1.\ee From the considerations of the
previous section, it is then obvious that if the above equalities are met, in any twisted sector
$|B\ket$, the representations $(S,V) \subset{\bf 27}$ and $(\bar S,V) \subset\overline{{\bf 27}}$
will be either simultaneously conserved or simultaneously destroyed, depending on the value of the
GGSO coefficients $[B|e_i],~[B|z_i]$. Explicitly building the spectrum and counting the states
surviving after the application of the various GGSO projections confirms the self-duality; we find
that a given twisted sector $|B\ket$ possesses one $SO(10)$-spinor (chiral or anti-chiral, its
chirality being fixed by the $\tilde b_2$-projection), one $SO(10)$-vector and one singlet under
$SO(10)$, but charged with respect to the additional $U(1)$ of $SO(10) \times U(1) \subset E_6$ :

\be |B\ket ~~~:~~~ (S,V)  \subset {\bf 27}~~~{\rm or}~~~(\bar S,V) \subset \overline{{\bf 27}}.\ee
When the action of all $z_i$-induced and $e_i$-induced freely-acting orbifolds are trivial on the
twisted sectors, we find therefore that the model possesses $N_+$ ${\bf 27}$ and $N_-$ $
\overline{{\bf 27}}$ $E_6$ representations, with $N_+ + N_-=48$. As the various orbifolds act, they
are able to cut in each twisted sector, either the vectorial, or the spinorial, or the whole
sector. As an example, we consider the twisted sectors
$|B^1_{\lambda_3\lambda_4\lambda_5\lambda_6}\ket$. Depending on the values of the GGSO coefficients
$[b_1|e_i]$ ,$i=1,2$ and $[e_j|e_i]$, $i=1,2$, $j=3,4,5,6$, we are able, thanks to the identities

\be \left[B^1_{\lambda_3\lambda_4\lambda_5\lambda_6}|e_i\right] = [B^1_{0000}|e_i]\prod_{j=3}^6
[e_j|e_i]^{\lambda_j},\ee

\be \left[B^1_{\lambda_3\lambda_4\lambda_5\lambda_6}|e_i\right] = [B^1_{0000}|z_i]\prod_{j=3}^6
[e_j|z_i]^{\lambda_j},\ee to determine the effect of the $e_i$- and $z_i$-projections on each one
of the twisted sectors of the $B^1$ plane. In particular, if $[e_k|e_i]=-1$, one sees that the
$e_i$-projection has opposite effects on the sectors
$\left|B^1_{\lambda_3\lambda_4\lambda_5\lambda_6}\right\ket$ and
$\left|B^1_{\lambda_3\lambda_4\lambda_5\lambda_6}+e_k\right\ket$.

\subsection{Duality inside the $\cN=2$ sectors}

As the classification in \cite{FKR,FKNR} shows, one can create several kinds of non-self-dual
models, in which, in a given twisted plane generated by the sectors $|B_1\ket$ and $|B_1+x\ket$,
one has either only spinorials of $SO(10)$ (with either positive or negative chirality; moreover,
the number of spinors and antispinors do not have to be equal) or only vectorials. For a non-self
dual model, as the $x$-superconformal map is broken, there exists at least one $i \in \{1\dots 6\}$
such that $[x|e_i]=-1$ or (inclusive) one $i \in \{1,2\}$ such that $[x|z_i]=-1$.

\bigskip

Let us start by considering a breaking by $e_i$. First we argue that the condition $[x|e_i]=-1$ is
able to break the self-duality only in the sectors where the freely-acting orbifold $e_i$ has the
possibility to project out entire representations of $SO(10)$ : namely $i=1,2$ for $B^1$ sectors,
$i=3,4$ for $B^2$ sectors, and $i=5,6$ for $B^3$ sectors. Indeed, let us suppose that $[x|e_1]=-1$
while the others $[x|e_i]=1$, and investigate the consequences on the spectrum. In the $B^1$
sectors, we have seen in a previous section that this breaking of $x$-map can project out spinors
and/or vectors of $SO(10)$. However, in $B^2$ and $B^3$ sectors, due to the intersections

\be \forall \lambda_i\in \{0,1\},~~~B^2_{\lambda_1\lambda_2\lambda_5\lambda_6}\cap e_1 =
(B^2_{\lambda_1\lambda_2\lambda_5\lambda_6}+x)\cap e_1 \neq \varnothing \ee and

\be \forall \lambda_i\in \{0,1\},~~~B^3_{\lambda_1\lambda_2\lambda_3\lambda_4}\cap e_1 =
(B^3_{\lambda_1\lambda_2\lambda_3\lambda_4}+x)\cap e_1 \neq \varnothing \ee the $e_1$-projection
only kills helicities, having a similar action in the sectors $B^{2,3}$ and their superconformal
partners $B^{2,3}+x$; it is not able to annihilate entire representations. Then the duality
spinor-vector is still valid in these sectors.

\bigskip

With this in mind, we focus on a case where the $x$-map is only broken in the first plane, that is
by $e_1$ and/or $e_2$. The duality map is then the following : \emph{the $(S_t\leftrightarrow
V)$-dual of a model where the $x$-map is broken only in the first twisted plane is constructed by
reversing the signs of the discrete torsion coefficients $[B^1_{0000}|e_i]$ and $[B^1_{0000}|z_j]$
for every $e_i$, $i=1,2$, satisfying $[x|e_i]=-1$, and for every $z_j$ satisfying $[x|z_j]=-1$.}
This procedure is easily seen to be in agreement with the rules given in \cite{FKR}, where the
general form of the duality transformation is formulated as the exchange of the ranks of the
matrices $\left[\Delta^{(1)},Y_{16}^{(1)}\right]$ and $\left[\Delta^{(1)},Y_{10}^{(1)}\right]$;
this particular set of rules actually exchanges the vectors $Y_{16}^{(1)}$ and $Y_{10}^{(1)}$.

\bigskip

\noindent   To prove this, let us suppose that $[x|e_1]=-1$ and consider the action of the $e_1$
projection on a given sector $\left|B^1_{\lambda_3\lambda_4\lambda_5\lambda_6}\right\ket$.

\bi \ittem Since one has

\be [B^1_{\lambda_3\lambda_4\lambda_5\lambda_6}|e_1] = [B^1_{0000}|e_1]\times
\underbrace{[e_3|e_1]^{\lambda_3}[e_4|e_1]^{\lambda_4}[e_5|e_1]^{\lambda_5}[e_6|e_1]^{\lambda_6}}_{=
\varepsilon} \ee we conclude that the sector $B^1_{\lambda_3\lambda_4\lambda_5\lambda_6}$ survives
the $e_1$ projection iff $[B^1_{0000}|e_i]=-\varepsilon$, and is projected out iff
$[B^1_{0000}|e_i]=\varepsilon$;

\ittem Then, since $[x|e_1]=-1$, we see that the sector
$B^1_{\lambda_3\lambda_4\lambda_5\lambda_6}+x$ survives iff $[B^1_{0000}|e_i]=\varepsilon$, and is
projected out iff $[B^1_{0000}|e_i]=-\varepsilon$. \ittem Therefore, the case
$[B^1_{0000}|e_i]=\varepsilon$ corresponds to keeping only the spinorial of $SO(10)$ arising from
$B^1_{\lambda_3\lambda_4\lambda_5\lambda_6}$, whereas $[B^1_{0000}|e_i]=-\varepsilon$ preserves
only the vectorial representation from this sector. \ittem Then, it is obvious to see that
reversing the sign of $[B^1_{0000}|e_1]$ will bring the dual model, since the factor
$\varepsilon=[e_3|e_1]^{\lambda_3}[e_4|e_1]^{\lambda_4}[e_5|e_1]^{\lambda_5}[e_6|e_1]^{\lambda_6}$
has not been changed in the process.

\ei

\bigskip

\noindent One must also look at the case where both $e_1$ and $e_2$ are breaking the $x$-map. It is
easy to convince oneself that one must reverse \emph{the two} discrete torsions $[B^1_{0000}|e_1]$
and $[B^1_{0000}|e_2]$ to get the dual model. Indeed, supposing that we start from a configuration
where only the spinorial representation survive from the sector
$B^1_{\lambda_3\lambda_4\lambda_5\lambda_6}$ after the two projections, one sees that reversing
only one of the two GGSO coefficients annihilates the whole sector
$B^1_{\lambda_3\lambda_4\lambda_5\lambda_6}$; whereas reversing both coefficients brings back the
vectorial of the sector.

\bigskip

Using similar arguments, one shows that, in the case of a breaking of the $x$-map by a set $z_i$,
the dual model is obtained by also switching the sign of the corresponding GGSO coefficient
$[B^1_{0000}|z_j]$. Indeed, the $z_i$ are never, in all three planes, part of the spin-fields
giving the vacuum, and then we can derive rules for them which are similar to the rules we have for
$e_{1,2}$ when acting on the first plane, $e_{3,4}$ on the second plane and $e_{5,6}$ on the third
plane. We note that, since the coefficients $(S|e_i)$ and $(S|z_i)$ are set to preserve $\cN=1$
supersymmetry, we may replace in the above rules $[B^1_{0000}|\dots]$ by $[b_1|\dots]$. We recover
the fact that the spinor-vector duality is realized \emph{within} each $\cN=2$ twisted plane
$B^{1,2,3}$.

\bigskip

Note that the rule we gave for the duality is not unique. One can check that, if we perform the duality
in the first plane, a dual model can be obtained by reversing the sign of $[B^1_{0000}|e_i]$ for every
$i$, $i=1\dots 6$, satisfying $[x|e_i]=-1$ (that is, we do not restrain ourselves to the two ``relevant''
projections in the first twisted plane which are $e_1$ and $e_2$). As a consequence, a given model admits
more than one dual. We will give additional arguments to this point at the end of this section.

\bigskip

When the $x$-map is broken in more than one plane, some subtleties arise, that require finer
details. Consider a $x$-map-breaking set $\alpha$, that is, $[\alpha|x]=-1$. $\alpha$ may be one of
the $e_i$ or one of the $z_i$. The duality operation has to be carried out in the three planes, by
reversing the GGSO coefficients $[b_1|\alpha]$, $[b_2|\alpha]$, and $[b_3|\alpha]$. However, the
third twisted plane is not independent from the two others, since $b_3=b_1+b_2+x$. Having carried
out the two first steps of the duality, we see that the two reversals \be \label{duality_2_planes}
[b_1|\alpha] \to -[b_1|\alpha], ~~~~[b_2|\alpha] \to -[b_2|\alpha] \ee entail, since
$[b_3|\alpha]=[b_1|\alpha]\cdot [b_2|\alpha]\cdot [x|\alpha]$ : \be [b_3|\alpha] \to [b_3|\alpha].
\ee This situation arises if a set $\alpha$ is able to break the spinor-vector duality in all three
planes. This is not the case for the $e_i$'s : as we have seen, $e_1$ and $e_2$ can only break the
duality in the first plane $B^1$, $e_3$ and $e_4$ in the second plane $B^2$, and
$e_5$ and $e_6$ in the third plane $B^3$. \\
It is however problematic when $\alpha$ is equal to $z_1$ and $z_2$. In that case, the duality is
restored if we assume the existence of $e_i$ and $e_j$, $i\neq j$, such as : \be
\label{additional_hypothesis} [e_i|z_1]=-1~~{\rm and}~~[e_j|z_2]=-1.\ee These conditions are
precisely the no-enhancements hypotheses we assumed to define the class of models in which we
demonstrate the duality.

\bigskip

\noindent Indeed, when \eqref{additional_hypothesis} is verified, the transformation
\eqref{duality_2_planes} for $\alpha=z_1$ entails\footnote{We suppose here that $e_i\neq e_5,e_6$.
If not, one adapts the proof in the straightforward way by exchanging the roles of $b_1,b_2,b_3$.}

\be \label{duality_3rd_plane} [b_3|z_1] \to -[b_3+e_i|z_1]. \ee This feature has the following
effect. In the two first twisted planes, the transformations \eqref{duality_2_planes} imply that
if, in a model, the sector $|B^1_{\lambda_3\lambda_4\lambda_5\lambda_6}\ket$ contains a spinorial
representation, it will contain a vectorial representation in the dual model. However, due to the
transformation \eqref{duality_3rd_plane}, we learn that if, in a model, the sector
$|B^3_{\lambda_1\lambda_2\lambda_3\lambda_4}\ket$ contains a spinorial representation, \emph{the
sector $|B^3_{\lambda_1\lambda_2\lambda_3\lambda_4}+e_i\ket$} will contain a vectorial
representation in the dual model. Then, in the third plane, we have a modified the $x$-map :
instead of linking a sector $|B^3_{\lambda_1\lambda_2\lambda_3\lambda_4}\ket$ to
$|B^3_{\lambda_1\lambda_2\lambda_3\lambda_4} + x\ket$, we have linked it to
$|B^3_{\lambda_1\lambda_2\lambda_3\lambda_4} + (x+e_i)\ket$. In this respect, the duality in the
third plane can also be viewed as being a sector-by-sector correspondence.

\bigskip

This also points out that the duality operation is not unique : one can choose to modify the $x$-map
$\alpha \mapsto \alpha +x$ into $\alpha \mapsto \alpha + x+e_i$ in the two first planes, for appropriate sets
$e_i$, \emph{i.e.} such as $\alpha + e_i$ is massless, and $e_i$ satisfies a condition of the type
\eqref{additional_hypothesis}. This observation is connected to the fact that the duality operation
is viewed in \cite{FKR} as an exchange of the rank of the matrices

\be {\rm rank}\left[\Delta^{(I)},Y_{16}^{(I)}\right]\leftrightarrow{\rm rank}
\left[\Delta^{(I)},Y_{10}^{(I)}\right]~;\ee this rank being constant under linear combinations on
the columns of $\Delta^{(I)}$.

\bigskip

Also note that when we will detail in section \ref{cocycle_insertions} the duality procedure, in
the no-enhancement framework, in terms of cocycle insertions, it will be sufficient to insert
cocycles relative to the twist parameters $h_1$ and $h_2$; the effect on the third plane will
automatically follow.

\bigskip

\subsection{Explicit realization of the duality in the first twisted plane}

We consider a model given by the following discrete torsion coefficients :

\be \label{coefficients} [B^1_{0000}|e_1]=1,~~~~~[x|e_1]=1,~~~~~[x|e_2]=-1; \ee and

\begin{center}
\begin{tabular}{|c|rr|}
\hline $[.|.]$ & $e_1$ & $e_2$ \\ \hline
$e_3$ & $-1$ & $1$ \\
$e_4$ & $1$ & $-1$ \\
$e_5$ & $1$ & $1$ \\
$e_6$ & $-1$ & $1$\\\hline
\end{tabular}
\end{center}
Then the action of $e_1$ and $e_2$ projections on the $B^1$ twisted plane and the resulting
spectrum are summarized in table \ref{BigTable}. This table gives, for a model and its dual, the
discrete torsion accounting for the effect of the projections $e_1$ and $e_2$ for each of the 16
sectors of the first twisted plane, and the corresponding surviving representations. The left part
of the table assumes $[B^1_{0000}|e_2]=1$ while the right part is for $[B^1_{0000}|e_2]=-1$. As we
discussed, a coefficient 1 relatively to $e_1$ projects out spinors and vectors altogether; a
coefficient 1 with respect to $e_2$ projects out spinors and a $-1$ projects out vectors.

\bigskip

\begin{table}[h!]
\begin{center}
\begin{tabular}{|c|rr|c||||rr|c|}
\hline $[.|.]$ & $e_1$ & $e_2$ & rep.& $e_1$ & $e_2$ & rep.\\ \hline
$B^1_{0000}$ & $1$ & $1$ & $\varnothing$ & $1$ & $-1$ & $\varnothing$\\
$B^1_{0001}$ & $-1$ & $1$ & $V$ & $-1$ & $-1$ & $S$\\
$B^1_{0010}$ & $1$ & $1$& $\varnothing$ & $1$ & $-1$& $\varnothing$ \\
$B^1_{0100}$ & $1$ & $-1$ & $\varnothing$ & $1$ & $1$ & $\varnothing$\\
$B^1_{1000}$ & $-1$ & $1$ & $V$ & $-1$ & $-1$ & $S$\\
$B^1_{1100}$ & $-1$ & $-1$ & $S$ & $-1$ & $1$ & $V$\\
$B^1_{1010}$ & $-1$ & $1$ & $V$ & $-1$ & $-1$ & $S$\\
$B^1_{1001}$ & $1$ & $1$ & $\varnothing$  & $1$ & $-1$ & $\varnothing$ \\
$B^1_{0101}$ & $-1$ & $-1$ & $S$ & $-1$ & $1$ & $V$\\
$B^1_{0110}$ & $1$ & $-1$ & $\varnothing$ & $1$ & $1$ & $\varnothing$  \\
$B^1_{0011}$ & $-1$ & $1$ & $V$ & $-1$ & $-1$ & $S$\\
$B^1_{1110}$ & $-1$ & $-1$ & $S$ & $-1$ & $1$ & $V$\\
$B^1_{1101}$ & $1$ & $-1$ & $\varnothing$  & $1$ & $1$ & $\varnothing$\\
$B^1_{1011}$ & $1$ & $1$ & $\varnothing$ & $1$ & $-1$ & $\varnothing$  \\
$B^1_{0111}$ & $-1$ & $-1$ & $S$ & $-1$ & $1$ & $V$\\
$B^1_{1111}$ & $1$ & $-1$ & $\varnothing$ & $1$ & $1$ & $\varnothing$  \\\hline
\end{tabular}
\end{center}
\caption{GGSO coefficients for the first twisted plane and corresponding surviving representation,
for the choice of coefficients \eqref{coefficients}.}\label{BigTable}
\end{table}

Note that in fact, this model is already self-dual; however, the duality operation is non-trivial,
as it exchanges spinorial and vectorial representations inside each twisted sector
$B^1_{\lambda_3\lambda_4\lambda_5\lambda_6}$, and we find it more instructive to detail the duality
procedure in this model rather than in a purely vectorial or purely spinorial model (recall from
\cite{FKR} that in one twisted plane, one has either a purely vectorial, purely
spinorial/anti-spinorial or half-vectorial half-spinorial -- \emph{i.e.} self-dual -- spectrum).
Obviously, under a duality transformation, a model having only spinorial representations (which can
be specifically obtained, for example, by setting $[e_{3,4,5,6}|e_2]=1$) will be related to a model
having only vectorial representations, the transformation being done sector by sector. We present
an explicit example of such a duality transformation in Appendix I.

\bigskip

We have not mentioned here the chirality of the spinorial representations; these depend on the
$\tilde b_2$ projection, which in turn depends on the discrete torsions

\be [B^1_{0000}|\tilde b^2];~~[e_i|\tilde b_2],~~i=3,4,5,6.\ee We will fix $[B^1_{0000}|\tilde
b^2]=-1$ and consider two cases of figure for the other four GGSO coefficients~:

\begin{center}

\begin{tabular}{m{3cm}m{2cm}m{3cm}}
(1)~:~~~\begin{tabular}{|c|r|} \hline $[.|.]$ & $\tilde b_2$ \\ \hline
$e_3$ & $1$\\
$e_4$ & $1$ \\
$e_5$ & $1$\\
$e_6$ & $1$\\\hline
\end{tabular} &
{\rm ~~~~~and} & (2)~:~~~
\begin{tabular}{|c|r|}
\hline $[.|.]$ & $\tilde b_2$ \\ \hline
$e_3$ & $1$\\
$e_4$ & $-1$ \\
$e_5$ & $1$\\
$e_6$ & $-1$\\\hline
\end{tabular}
\end{tabular}
\end{center}

\noindent Extracting the spinorial representations from the previous model, we find that for case
$(1)$, before and after duality, all $SO(10)$ spinors have positive chirality. For case $(2)$, we
find that, before and after duality, we have 2 chiral and 2 anti-chiral spinors.

\bigskip

\noindent Note that to put in evidence more features of the construction, we have taken non-trivial
values for the coefficients $[e_{3,4,5,6}|e_{1,2}]$. Had we not done this, the remaining model
would have had more generations. One sees that within a twisted plane, arbitrary values of the
coefficients $[e_{3,4,5,6}|e_i]$, where $e_i$ doesn't break the $x$-map, are only able to project
out half of the twisted sectors; only 8 sectors out of 16 contribute, giving either a purely
spinorial, purely vectorial, or half-vectorial and half-spinorial spectrum.

\bigskip

Further projections in the spectrum can then be performed by acting with the orbifolds generated by
$z_1$ and $z_2$. Indeed we can obtain the formula

\be [B^1_{\lambda_3\lambda_4\lambda_5\lambda_6}|z_{1,2}] = [B^1_{0000}|z_{1,2}]\times \prod_{i=3}^6
[e_i|z_{1,2}]^{\lambda_i}\ee and the survival condition of the sector
$\left|B^1_{\lambda_3\lambda_4\lambda_5\lambda_6}\right\ket$ is

\be [B^1_{\lambda_3\lambda_4\lambda_5\lambda_6}|z_{1,2}] = -1. \ee Setting, for some $(i,j) \in
\{3,4,5,6\} \times \{1,2\}$, some discrete torsions

\be [e_i|z_j] = -1 \ee gives one access to models in which only 4 sectors or only 2 sectors out of
the 16 survive at the massless level.

\bigskip

To conclude this subsection, let us note that the explicit model we constructed above is self-dual;
however the $E_6$ gauge symmetry has been broken. Breaking $E_6 \to SO(10) \times U(1)$, as it
makes an abelian factor $U(1)$ appear in the gauge group, is generally believed to lead to
anomalies. However, in a class of self-dual models, $U(1)$ anomalies can be evaded  when summing on
the contribution of the three twisted planes. We provide an explicit example of this property in
the Appendix II.

\subsection{Plane by plane insertions of discrete torsion coefficients, and their overall effects}
\label{cocycle_insertions}

In this subsection, we want to indicate how these constructions can be translated in terms of
modifications of the overall phase $\Phi$ introduced in the general form of the partition function
\eqref{partition_function}. Again, we focus on the first twisted plane; the generalization for the
simultaneous action on the three planes will be addressed at the end of this subsection. We are
then considering the internal dimensions $e_i$, $i=1,2$. The term of the partition function
representing the first twisted plane is obtained when the four space-time fermions $\chi^{3,4,5,6}$
are twisted : therefore,
$h_1=0$ and, $h_2=h_3=h$ is the relevant twisting parameter.\\

Remembering that the freely-acting orbifolds are conveniently represented by the
insertion of cocycles in the partition function, we find the following rules.\\
First, in the absence of superconformal symmetry breaking, one is able to project out a whole
sector of the twisted plane (that is, both the spinorial and the vectorial coming from this sector)
by adding a phase

\be \label{all_sectors_out} (-)^{h t_i + g s_i}~,~~~~(-)^{h G_i + g H_i}~, \ee depending on the
breaking being done by a $e_i$ or a $z_i$ projection. As we discussed earlier, such a coupling
renders the $h=1$ sectors massive, which is the case in the plane that we are considering.
Furthermore, as we have explained before, the effect the different sectors of the plane is dictated
by the values of the coefficients $(e_i|e_j)$. These discrete torsions are controlled by the
insertion of the cocycles

\be (-)^{s_i t_j + s_j t_i}. \ee One is then able to construct a variety of self-dual models using
these rules. Similarly, one is able to control the value of the coefficient $(e_i|z_j)$ by means of
the insertion of

\be (-)^{s_i G_j + H_j t_i}. \ee


\bigskip

\noindent  The superconformal $x$-map is broken as soon as we couple a freely-acting orbifold to
the $SO(10)$ spin-structure $(\epsilon,\xi)$. In the first twisted plane, such a breaking requires
the action of at least one of the sets $(e_1,e_2,z_1,z_2)$; the corresponding cocycles to be
inserted then read, respectively :

\be \label{x_breaking_cocycle} (-)^{\epsilon t_i + \xi s_i + s_i t_i},~~i=1,2~;~~~~ (-)^{\epsilon
G_i + \xi H_i + H_i G_i},~~i=1,2~.  \ee Coupling the two previous effects now allow us to control
which representation (spinorial or vectorial) survives at the massless level in the model. Starting
from a case where both spinors and vectors survive, the addition of one of the SCFT-breaking phases
\eqref{x_breaking_cocycle} lifts the spinorials of $SO(10)$, so that only the vectorials survive.
If instead we start from a case  where the whole sector has been projected out, by the insertion of
a cocycle of the form \eqref{all_sectors_out}, adding a cocycle \eqref{x_breaking_cocycle} recovers
the spinorials, while the vectorials remain massive. The phase we inserted in this case is then the
product of
\eqref{all_sectors_out} and \eqref{x_breaking_cocycle}.\\
We may summarize the possibilities as follows :

\bi \item no cocycle introduced : $S$ and $V$ stay at the massless level;

\item $(-)^{h t_i + g s_i}$ : both $S$ and $V$ become massive;

\item $(-)^{\epsilon t_i + \xi s_i + s_i t_i}$ : $S$ becomes massive, $V$ stays massless;

\item $(-)^{(\epsilon + h) t_i + (\xi+g) s_i + s_i t_i}$ : $S$ stays massless, $V$
becomes massive.

\ei Of course, if one considers a breaking by $z_i$, one has to replace $(s_i,t_i)$ by $(H_i,G_i)$.

\bigskip

We then learn how to engineer the duality map directly on the partition function. We have stated
that it has to be done by reversing the GGSO projections $[B^1|e_i]$, $[B^1|z_i]$ for each
$x$-breaking projections $e_i$, $z_i$. But these values are encoded in cocycles \be (-)^{h t_i + g
s_i}~,~~~~(-)^{h G_i + g H_i}~. \ee where $h$ is the orbifold parameter relevant for the plane we
are interested in. Therefore, to carry out the duality map, one has to insert a cocycle
\eqref{all_sectors_out} for each projection breaking the $x$-map (\emph{i.e.} such that a cocycle
of the form \eqref{x_breaking_cocycle} is present in the partition function).

\section{Conclusion and discussion}

In this paper, we gave a new demonstration of the spinor-vector duality that was shown to hold
among the $\cN=2$ $\Z_2$ and the $\cN=1$, $\Z_2\times \Z_2$ heterotic--string vacua obtained
\emph{via} the free fermionic construction. We interpreted the freely-acting orbifolds present in
the model in terms of stringy Scherk-Schwarz mechanisms; these have been used to give a
non-vanishing mass to some sectors of the theory, and/or to perform a spontaneous breaking of the
right-moving superconformal algebra (also called $x$-map) which is responsible of the gauge
enhancement $SO(10) \times U(1) \to E_6$. Such a breaking creates non-self-dual models, where we do
not have the same number of spinorial and vectorial representations of $SO(10)$ at the massless
level of the theory. We described the procedure used to construct the dual of a given model.
Moreover, we explicitly constructed self-dual models in which $E_6$ gauge is broken.

Such models may, or may not, be free from all Abelian and mixed anomalies. The cases in which the
self--dual models are particularly interested, as in such models one does not need to resort to
field theory arguments to shift the vacuum to a stable supersymmetric vacuum. Finally, we have
given rules on how to perform this duality directly on the expression of the 1-loop partition
function of the model.

One may ask what are the implications of such a duality. Firstly, we can see it as a symmetry in
the space of vacua of string theory, whose study has been of great interest over the past years
\cite{landscape}. Furthermore, the duality is exhibited in the space of free fermionic models that
have also given rise to some of the most realistic string models constructed to date. The
geometrical structure underlying the free fermionic models is that of the $\Z_2\times \Z_2$
orbifold, and a natural question is whether it extends to other orbifolds. The spinor--vector
duality can be thought of as being of the same kind as mirror symmetry \cite{mirror}. Indeed,
mirror symmetry is manifest in this model as the symmetry exchanging spinorials of $SO(10)$ into
anti-spinorials of $SO(10)$. This is due to the Type II $\leftrightarrow$ Heterotic correspondence
being related to the embedding of the spin-connection in the gauge connection. Therefore, changing
the chirality of the $SO(10)$ spinors amounts, on the Type II side, to change the GSO projection on
the right-hand side of the theory. This Type IIA $\leftrightarrow$ Type IIB switch is known
\cite{StroYau} to be equivalent to the substitution of the compactification manifold by its mirror.
Our constructions displays this mirror symmetry : this relies on the choice of the coefficients
$[b_1|\tilde b_2]$ and $[e_i|\tilde b_2]$, as we have shown that the $\tilde b_2$ projection
imposes the chirality of the massless spinorial representations (if any). The mirror symmetry
implies a change in the topology of the compactification manifold, as the Euler characteristic is
taken to its opposite. Spinor-vector duality can, as well, be thought of as another
topology--changing duality. Note that its range of application is wider than the mirror case. Here,
non-self-dual points correspond to $N=(2,0)$ compactifications. Just as mirror symmetry can be
thought of as a manifestation of $T$--duality \cite{StroYau} also the spinor--vector duality may be
regarded as such, but with the added action on the bundle representing the gauge degrees of freedom
of the heterotic string, induced by the breaking of the $N=2$ world--sheet superconformal symmetry
on the right-moving bosonic side of the heterotic string. Thus, just as mirror symmetry have led to
the notion of topology changing transition between mirror manifolds, the spinor--vector duality
suggests that the web of connections is far more complex, and further demonstrating that our
understanding of string theory is truly only rudimentary. Furthermore, what we may find is that the
distinction of particles into spinor and vector representation is a mere low energy organisation.
What the string truly cares about is its internal consistency, characterized by the modular
invariance of the partition function.

\section*{Acknowledgements}

AEF and JR would like to thank the \'Ecole Normale Sup\'erieure and the \'Ecole Polytechnique for
hospitality. AEF work is supported in part by STFC under contract PP/D000416/1 and by the EU under
contract MRTN--CT--2006--035863--1. CK is supported in part by the EU under the contracts
MRTN-CT-2004-005104, MRTN-CT-2004-512194, MRTN-CT-2004-503369, MEXT-CT-2003-509661, ANR (CNRS-USAR)
contract 05-BLAN-0079-01. JR work is supported in part by the EU under contracts
MRTN--CT--2006--035863--1 and  MRTN--CT--2004--503369.

\newpage
\appendix
\renewcommand{\theequation}{I.\arabic{equation}}
\renewcommand{\thesection}{I.}
\setcounter{equation}{0}

\begin{center}
{\Large \bf Appendix I\\ ~\\A dual pair of models with spectrum in the first twisted plane}
\end{center}

We consider the model given by the following GGSO coefficients matrix :

\be [v_i|v_j] = e^{i\pi (v_i|v_j)} \ee

\be \label{BigMatrix}  (v_i|v_j)\ \ =\ \ \bordermatrix{ &1 &S
&e_1&e_2&e_3&e_4&e_5&e_6&b_1&b_2&z_1&z_2\cr
 1 & 1& 1& 1& 1& 1& 1& 1& 1& 1& 1& 1& 1\cr
 S & 1& 1& 1& 1& 1& 1& 1& 1& 1& 1& 1& 1\cr
e_1& 1& 1& 0& 1& 0& 0& 0& 1& 0& 0& 1& 1\cr 
e_2& 1& 1& 1& 0& 0& 0& 1& 1& 0& 0& 0& 0\cr 
e_3& 1& 1& 0& 0& 0& 0& 0& 0& 0& 0& 0& 0\cr 
e_4& 1& 1& 0& 0& 0& 0& 0& 0& 0& 1& 1& 1\cr 
e_5& 1& 1& 0& 1& 0& 0& 0& 0& 0& 0& 1& 0\cr 
e_6& 1& 1& 1& 1& 0& 0& 0& 0& 1& 1& 1& 0\cr 
b_1& 1& 0& 0& 0& 0& 0& 0& 1& 1& 0& 0& 1\cr 
b_2& 1& 0& 0& 0& 0& 1& 0& 1& 1& 1& 0& 0\cr 
z_1& 1& 1& 1& 0& 0& 1& 1& 1& 0& 0& 1& 0\cr 
z_2& 1& 1& 1& 0& 0& 1& 0& 0& 1& 0& 0& 1\cr
  }
\ee

As far as the $SO(10)$ representations are concerned, this model contains two vectorials ${\bf
10}$, one in the sector $S+b_1+e_5+x$, and one in the sector $S+b_1+e_3+e_5+x$. The spectrum is
therefore contained in the first twisted plane; we will only need to carry out the duality in this
plane.\\
We apply the duality procedure as follows.\\ First, we notice that, since $$x = 1+S+\sum_{i=1}^6
e_i+z_1+z_2,$$ we have

$$(x|e_1)=0,~~~~(x|e_2)=1,~~~~(x|e_3)=0,$$
$$(x|e_4)=0,~~~~(x|e_5)=0,~~~~(x|e_6)=1,$$
$$(x|z_1)=1,~~~~(x|z_2)=1. $$
The method we exposed then consists in reversing the GGSO coefficients $(b_1|e_2)$,
$(b_1|z_1)$ and $(b_1|z_2)$. The resulting matrix is therefore (the coefficients we changed are 
in bold) :

\be \label{BigMatrixDual}  (v_i|v_j)\ \ =\ \ \bordermatrix{ &1 &S
&e_1&e_2&e_3&e_4&e_5&e_6&b_1&b_2&z_1&z_2\cr
 1 & 1& 1& 1& 1& 1& 1& 1& 1& 1& 1& 1& 1\cr
 S & 1& 1& 1& 1& 1& 1& 1& 1& 1& 1& 1& 1\cr
e_1& 1& 1& 0& 1& 0& 0& 0& 1& 0& 0& 1& 1\cr 
e_2& 1& 1& 1& 0& 0& 0& 1& 1& {\bf 1}& 0& 0& 0\cr 
e_3& 1& 1& 0& 0& 0& 0& 0& 0& 0& 0& 0& 0\cr 
e_4& 1& 1& 0& 0& 0& 0& 0& 0& 0& 1& 1& 1\cr 
e_5& 1& 1& 0&1& 0& 0& 0& 0& 0& 0& 1& 0\cr 
e_6& 1& 1& 1& 1& 0& 0& 0& 0& 1& 1& 1& 0\cr 
b_1& 1& 0& 0& {\bf 1}& 0& 0& 0& 1& 1& 0& {\bf 1}& {\bf 0}\cr 
b_2& 1& 0& 0& 0& 0& 1& 0& 1& 1& 1& 0& 0\cr 
z_1& 1& 1& 1& 0& 0& 1& 1& 1& {\bf 1}& 0& 1& 0\cr 
z_2& 1& 1& 1& 0& 0& 1& 0& 0& {\bf 0}& 0& 0& 1\cr
}
\ee

When explicitly computing the spectrum of this new model, we find indeed that two spinors
${\overline{\bf 16}}$ of $SO(10)$ arise from the first plane, in the sectors $S+b_1+e_5$ and
$S+b_1+e_3+e_5$. We see then that in this simple case, the duality transformation occurs sector by
sector in the first twisted plane, like described in section \ref{construction}.

\appendix
\renewcommand{\theequation}{II.\arabic{equation}}
\renewcommand{\thesection}{II.}
\setcounter{equation}{0}

\begin{center}
{\Large \bf Appendix II\\ ~\\A self-dual, anomaly-free model without $E_6$ enhancement}
\end{center}


%
%
%

We are considering the model given by the matrix which coefficents $(v_i|v_j) \in \{0,1\}$ are
defined by

\be [v_i|v_j] = e^{i\pi (v_i|v_j)} \ee

\be \label{BigMatrixII}  (v_i|v_j)\ \ =\ \ \bordermatrix{ &1 &S
&e_1&e_2&e_3&e_4&e_5&e_6&b_1&b_2&z_1&z_2\cr
 1 & 1& 1& 1& 1& 1& 1& 1& 1& 1& 1& 1& 1\cr
 S & 1& 1& 1& 1& 1& 1& 1& 1& 1& 1& 1& 1\cr
e_1& 1& 1& 0& 0& 0& 0& 0& 1& 1& 0& 0& 1\cr 
e_2& 1& 1& 0& 0& 1& 1& 1& 1& 1& 0& 0& 0\cr 
e_3& 1& 1& 0& 1& 0& 1& 1& 0& 0& 0& 0& 0\cr
 e_4& 1& 1& 0& 1& 1& 0& 1& 1& 0& 1& 1& 0\cr
 e_5& 1& 1& 0& 1& 1& 1& 0& 1& 0& 0& 1& 1\cr 
e_6& 1& 1& 1& 1& 0& 1& 1& 0& 1& 0& 0& 1\cr
 b_1& 1& 0& 1& 1& 0& 0& 0& 1& 1& 0& 1& 1\cr
 b_2& 1& 0& 0& 0& 0& 1& 0& 0& 1& 1& 1& 0\cr
 z_1& 1& 1& 0& 0& 0& 1& 1& 0& 1& 1& 1& 0\cr 
z_2& 1& 1& 1& 0& 0& 0& 1& 1& 0& 0& 0& 1\cr
  }
\ee We see that since $(z_1|x) = (z_1|1)+(z_1|S)+\sum_{i=1}^6(z_1|e_i) + (z_1|z_1) + (z_1|z_2)
\equiv 1$ mod. 2, the gauge group $E_6$ is broken. Moreover, the conditions $(e_1|z_2)=(e_4|z_1)=1$
ensure that the ``hidden'' gauge group is minimal and the full gauge group is
$SO(10)\times{U(1)}^3\times{SO(8)}\times{SO(8)}$. The spectrum of this model contains (we note as
an index the three charges under the $U(1)_{\bar\eta^i}$, $i=1,2,3$) :

\bi
\item three spinors ${\bf 16}$ of $SO(10)$, one for each twisted plane,
$${\bf 16}_{(1/2,0,0)},~{\bf 16}_{(0,-1/2,0)},~{\bf 16}_{(0,0,-1/2)},$$
\item  three vectors ${\bf 10}$ of $SO(10)$, one for each twisted plane,
$${\bf 10}_{(0,1/2,1/2)},~{\bf 10}_{(-1/2,0,1/2)},~{\bf 10}_{(-1/2,1/2,0)},$$
\item six non-abelian gauge group singlets, two for each twisted plane,
$${\bf 1}_{(1 ,-1/2,-1/2)},~{\bf 1}_{(1/2,1,-1/2)},~{\bf 1}_{(1/2,-1/2,1)},$$
$${\bf 1}_{(-1 ,-1/2,-1/2)},~{\bf 1}_{(1/2,-1,-1/2)},~{\bf 1}_{(1/2,-1/2,-1)}.$$
\ei By verifying the identities $\sum q_i = \sum q_i^3 = 0$ for the three abelian factors of the
gauge group, we see that the observable spectrum is anomaly-free. Note that this anomaly does not
occur plane by plane, but results
from a cancellation between the three planes.\\
One can also check that in this model, the contributions of the $({\bf 8},\bf 1)$ and $(\bf 1,{\bf
8})$ multiplets of $SO(8) \times SO(8)$ to the $U(1)$ anomalies cancel.


\newpage


\begin{thebibliography}{99}





\bibitem{Gross}
  D.~J.~Gross, J.~A.~Harvey, E.~J.~Martinec and R.~Rohm,
  ``Heterotic String Theory. 1. The Free Heterotic String,''
  Nucl.\ Phys.\  B {\bf 256}, 253 (1985).\\
  D.~J.~Gross, J.~A.~Harvey, E.~J.~Martinec and R.~Rohm,
  ``Heterotic String Theory. 2. The Interacting Heterotic String,''
  Nucl.\ Phys.\  B {\bf 267}, 75 (1986).


\bibitem{CalabiYau}
  P.~Candelas, G.~T.~Horowitz, A.~Strominger and E.~Witten,
  ``Vacuum Configurations For Superstrings,''
  Nucl.\ Phys.\  B {\bf 258}, 46 (1985).


\bibitem{StringySS}
  C.~Kounnas and M.~Porrati,
  ``Spontaneous Supersymmetry Breaking in String Theory,''
  Nucl.\ Phys.\  B {\bf 310} (1988) 355.\\
  C.~Kounnas and B.~Rostand,
  ``Coordinate Dependent Compactifications and Discrete Symmetries,''
  Nucl.\ Phys.\  B {\bf 341}, 641 (1990).


\bibitem{FKR}
  A.~E.~Faraggi, C.~Kounnas and J.~Rizos,
  ``Chiral family classification of fermionic Z(2) x Z(2) heterotic orbifold
  models,''
  Phys.\ Lett.\  B {\bf 648} (2007) 84
  [arXiv:hep-th/0606144].\\
  A.~E.~Faraggi, C.~Kounnas and J.~Rizos,
  ``Spinor - vector duality in fermionic Z(2) x Z(2) heterotic orbifold
  models,''
  Nucl.\ Phys.\  B {\bf 774} (2007) 208
  [arXiv:hep-th/0611251].\\
  A.~E.~Faraggi, C.~Kounnas and J.~Rizos,
  ``Spinor-Vector Duality in N=2 Heterotic String Vacua,''
  Nucl.\ Phys.\  B {\bf 799} (2008) 19
  [arXiv:0712.0747 [hep-th]].

\bibitem{FF}
  I.~Antoniadis, C.~Bachas, C.~Kounnas and P.~Windey,
  ``Supersymmetry Among Free Fermions And Superstrings,''
  Phys.\ Lett.\  B {\bf 171}, 51 (1986).\\
  I.~Antoniadis, C.~P.~Bachas and C.~Kounnas,
  ``Four-Dimensional Superstrings,''
  Nucl.\ Phys.\  B {\bf 289}, 87 (1987).\\
  H.~Kawai, D.~C.~Lewellen and S.~H.~H.~Tye,
  ``Construction of Fermionic String Models in Four-Dimensions,''
  Nucl.\ Phys.\  B {\bf 288}, 1 (1987).\\
  I.~Antoniadis and C.~Bachas,
  ``4-D Fermionic Superstrings with Arbitrary Twists,''
  Nucl.\ Phys.\  B {\bf 298} (1988) 586.


\bibitem{Realistic}
  I.~Antoniadis, J.~R.~Ellis, J.~S.~Hagelin and D.~V.~Nanopoulos,
  ``The Flipped SU(5) x U(1) String Model Revamped,''
  Phys.\ Lett.\  B {\bf 231} (1989) 65.\\
  A.~E.~Faraggi, D.~V.~Nanopoulos and K.~j.~Yuan,
  ``A Standard Like Model in the 4D Free Fermionic String Formulation,''
  Nucl.\ Phys.\  B {\bf 335}, 347 (1990).\\
  I.~Antoniadis, G.~K.~Leontaris and J.~Rizos,
  ``A Three generation SU(4) x O(4) string model,''
  Phys.\ Lett.\  B {\bf 245}, 161 (1990).\\
    A.~E.~Faraggi and D.~V.~Nanopoulos,
``Naturalness of three generations in free fermionic Z(2)-n x Z(4) string models''
Phys.\ Rev.\ D {\bf 48}, 3288 (1993).\\
    A.~E.~Faraggi,
  ``A New standard - like model in the four-dimensional free fermionic string
 formulation,''
 Phys.\ Lett.\  B {\bf 278}, 131 (1992)\\
  G.~K.~Leontaris and J.~Rizos,
  ``N = 1 supersymmetric SU(4) x SU(2)L x SU(2)R effective theory from the
  weakly coupled heterotic superstring,''
  Nucl.\ Phys.\  B {\bf 554}, 3 (1999)
  [arXiv:hep-th/9901098].\\
  G.~B.~Cleaver, A.~E.~Faraggi and D.~V.~Nanopoulos,
  ``String derived MSSM and M-theory unification,''
  Phys.\ Lett.\  B {\bf 455}, 135 (1999)
  [arXiv:hep-ph/9811427].\\
  G.~B.~Cleaver, A.~E.~Faraggi and C.~Savage,
 ``Left-right symmetric heterotic-string derived models,''
  Phys.\ Rev.\  D {\bf 63}, 066001 (2001)
  [arXiv:hep-ph/0006331].\\
  A.~E.~Faraggi, E.~Manno and C.~Timirgaziu,
  ``Minimal standard heterotic string models,''
  Eur.\ Phys.\ J.\  C {\bf 50}, 701 (2007)
  [arXiv:hep-th/0610118].



\bibitem{z2z2models1}
  A.~E.~Faraggi,
  ``Z(2) x Z(2) Orbifold compactification as the origin of realistic free
  fermionic models,''
  Phys.\ Lett.\  B {\bf 326}, 62 (1994)
  [arXiv:hep-ph/9311312].\\
  P.~Berglund, J.~R.~Ellis, A.~E.~Faraggi, D.~V.~Nanopoulos and Z.~Qiu,
  ``Elevating the free-fermion Z(2) x Z(2) orbifold model to a
  compactification of F-theory,''
  Int.\ J.\ Mod.\ Phys.\  A {\bf 15}, 1345 (2000)
  [arXiv:hep-th/9812141].\\
  R.~Donagi and A.~E.~Faraggi,
  ``On the number of chiral generations in Z(2) x Z(2) orbifolds,''
  Nucl.\ Phys.\  B {\bf 694}, 187 (2004)
  [arXiv:hep-th/0403272].\\



\bibitem{z2z2models2}
  E.~Kiritsis, C.~Kounnas, P.~M.~Petropoulos and J.~Rizos,
  ``On the Heterotic Effective Action at One-Loop, Gauge Couplings and the
  Gravitational Sector,''
  arXiv:hep-th/9605011.\\
  E.~Kiritsis, C.~Kounnas, P.~M.~Petropoulos and J.~Rizos,
  ``Universality properties of N = 2 and N = 1 heterotic threshold
  corrections,''
  Nucl.\ Phys.\  B {\bf 483}, 141 (1997)
  [arXiv:hep-th/9608034].\\
  E.~Kiritsis and C.~Kounnas,
  ``Perturbative and non-perturbative partial supersymmetry breaking:  N = 4
  --> N = 2 --> N = 1,''
  Nucl.\ Phys.\  B {\bf 503}, 117 (1997)
  [arXiv:hep-th/9703059].\\
  A.~Gregori and C.~Kounnas,
  ``Four-dimensional N = 2 superstring constructions and their
  (non-)perturbative duality connections,''
  Nucl.\ Phys.\  B {\bf 560}, 135 (1999)
  [arXiv:hep-th/9904151].\\
    A.~Gregori, C.~Kounnas and J.~Rizos,
  ``Classification of the N = 2, Z(2) x Z(2)-symmetric type II orbifolds  and
  their type II asymmetric duals,''
  Nucl.\ Phys.\  B {\bf 549}, 16 (1999)
  [arXiv:hep-th/9901123].




\bibitem{FKNR}
  A.~E.~Faraggi, C.~Kounnas, S.~E.~M.~Nooij and J.~Rizos,
  ``Classification of the chiral Z(2) x Z(2) fermionic models in the  heterotic
  superstring,''
  Nucl.\ Phys.\  B {\bf 695} (2004) 41
  [arXiv:hep-th/0403058].



\bibitem{SS}
  J.~Scherk and J.~H.~Schwarz,
  ``Spontaneous Breaking Of Supersymmetry Through Dimensional Reduction,''
  Phys.\ Lett.\  B {\bf 82}, 60 (1979).\\
  J.~Scherk and J.~H.~Schwarz,
  ``How To Get Masses From Extra Dimensions,''
  Nucl.\ Phys.\  B {\bf 153}, 61 (1979).

\bibitem{CJKPT}
  T.~Catelin-Jullien, C.~Kounnas, H.~Partouche and N.~Toumbas,
  ``Thermal/quantum effects and induced superstring cosmologies,''
  Nucl.\ Phys.\  B {\bf 797}, 137 (2008)
  [arXiv:0710.3895 [hep-th]].

\bibitem{Nooij}
  S.~E.~M.~Nooij,
  ``Classification of the chiral Z(2) x Z(2) heterotic string models,''
  arXiv:hep-th/0603035.



\bibitem{GepnerLust}
  D.~Gepner,
  ``Space-Time Supersymmetry in Compactified String Theory and Superconformal
  Models,''
  Nucl.\ Phys.\  B {\bf 296}, 757 (1988).\\
  W.~Lerche, D.~Lust and A.~N.~Schellekens,
  ``Chiral Four-Dimensional Heterotic Strings from Selfdual Lattices,''
  Nucl.\ Phys.\  B {\bf 287}, 477 (1987).

\bibitem{GreenSchwarz}
  M.~B.~Green and J.~H.~Schwarz,
  ``Anomaly Cancellation In Supersymmetric D=10 Gauge Theory And Superstring
  Theory,''
  Phys.\ Lett.\  B {\bf 149}, 117 (1984).


\bibitem{xmap}
  A.~E.~Faraggi,
  ``Generation mass hierarchy in superstring derived models,''
  Nucl.\ Phys.\  B {\bf 407}, 57 (1993)
  [arXiv:hep-ph/9210256].\\
  A.~E.~Faraggi,
  ``Partition functions of NAHE-based free fermionic string models,''
  Phys.\ Lett.\  B {\bf 544}, 207 (2002)
  [arXiv:hep-th/0206165].

\bibitem{landscape}
  D.~Senechal,
  ``Search For Four-Dimensional String Models. 1,''
  Phys.\ Rev.\  D {\bf 39}, 3717 (1989).\\
  K.~R.~Dienes,
  ``New string partition functions with vanishing cosmological constant,''
  Phys.\ Rev.\ Lett.\  {\bf 65}, 1979 (1990).\\
  M.~R.~Douglas,
  ``The statistics of string / M theory vacua,''
  JHEP {\bf 0305}, 046 (2003)
  [arXiv:hep-th/0303194].\\
  F.~Denef and M.~R.~Douglas,
  ``Distributions of flux vacua,''
  JHEP {\bf 0405}, 072 (2004)
  [arXiv:hep-th/0404116].\\
  B.~S.~Acharya, F.~Denef and R.~Valandro,
 ``Statistics of M theory vacua,''
  JHEP {\bf 0506}, 056 (2005)
  [arXiv:hep-th/0502060].\\
  R.~Blumenhagen, F.~Gmeiner, G.~Honecker, D.~Lust and T.~Weigand,
  ``The statistics of supersymmetric D-brane models,''
  Nucl.\ Phys.\  B {\bf 713}, 83 (2005)
  [arXiv:hep-th/0411173].\\
   K.~R.~Dienes,
  ``Statistics on the heterotic landscape: Gauge groups and cosmological
  constants of four-dimensional heterotic strings,''
  Phys.\ Rev.\  D {\bf 73}, 106010 (2006)
  [arXiv:hep-th/0602286].\\
  M.~R.~Douglas and W.~Taylor,
  ``The landscape of intersecting brane models,''
  JHEP {\bf 0701}, 031 (2007)
  [arXiv:hep-th/0606109].\\
  K.~R.~Dienes, M.~Lennek, D.~Senechal and V.~Wasnik,
  ``Supersymmetry versus Gauge Symmetry on the Heterotic Landscape,''
  Phys.\ Rev.\  D {\bf 75}, 126005 (2007)
  [arXiv:0704.1320 [hep-th]].\\
  O.~Lebedev, H.~P.~Nilles, S.~Raby, S.~Ramos-Sanchez, M.~Ratz, P.~K.~S.~Vaudrevange and A.~Wingerter,
  ``A mini-landscape of exact MSSM spectra in heterotic orbifolds,''
  Phys.\ Lett.\  B {\bf 645}, 88 (2007)
  [arXiv:hep-th/0611095].


\bibitem{mirror}
  B.~R.~Greene and M.~R.~Plesser,
  ``Duality In Calabi-Yau Moduli Space,''
  Nucl.\ Phys.\  B {\bf 338}, 15 (1990).


\bibitem{StroYau}
  A.~Strominger, S.~T.~Yau and E.~Zaslow,
  ``Mirror symmetry is T-duality,''
  Nucl.\ Phys.\  B {\bf 479}, 243 (1996)
  [arXiv:hep-th/9606040].











\end{thebibliography}
\end{document}